\documentclass[manuscript]{aastex}
\usepackage{amsmath}
\usepackage{comment}
\usepackage{grffile}
\usepackage{lscape}
\usepackage{arrayjobx}
\usepackage{multirow}
\usepackage{fancybox}

\newcommand{\iffigure}{\iftrue}

\newcommand{\iffigurePV}{\iftrue}

\newcommand{\dirfig}{}
\newcommand{\dirfigchan}{}

\newcommand{\bff}{}

\newcommand{\CO}{C$^{17}$O}
\newcommand{\co}{$J$ = 2--1}
\newcommand{\TFA}{H$_2$CS}
\newcommand{\tfaa}{$7_{0, 7}$--$6_{0, 6}$}
\newcommand{\tfab}{$7_{2, 5}$--$6_{2, 4}$}
\newcommand{\tfac}{$7_{2, 6}$--$6_{2, 5}$}
\newcommand{\tfada}{$7_{4, 3}$--$6_{4, 2}$}
\newcommand{\tfadb}{$7_{4, 4}$--$6_{4, 3}$}
\newcommand{\tfad}{\tfada, \tfadb}
\newcommand{\tfaea}{$7_{3, 4}$--$6_{3, 3}$}
\newcommand{\tfaeb}{$7_{3, 5}$--$6_{3, 4}$}
\newcommand{\tfae}{\tfaea, \tfaeb}
\newcommand{\tfafa}{$7_{5, 2}$--$6_{5, 1}$}
\newcommand{\tfafb}{$7_{5, 3}$--$6_{5, 2}$}
\newcommand{\tfaf}{\tfafa, \tfafb} 

\newcommand{\MN}{CH$_3$OH}
\newcommand{\MF}{HCOOCH$_3$}
\newcommand{\EC}{C$_2$H$_5$CN}
\newcommand{\ec}{$28_{1, 28}$--$27_{1, 27}$}

\newcommand{\ocs}{$J$ = 19--18}

\newcommand{\iras}{IRAS 16293--2422}

\newcommand{\ire}{infalling-rotating envelope}

\newcommand{\cb}{centrifugal barrier} 

\newcommand{\desys}{disk/envelope system}
\newcommand{\dfr}{disk-forming region}
\newcommand{\ia}{inclination angle}
\newcommand{\incRem}{0\degr\ for a face-on configuration}
\newcommand{\los}{line of sight}

\newcommand{\sysV}{systemic velocity}
\newcommand{\Eu}{$E_{\rm u}$}

\newcommand{\Msun}{$M_\odot$}

\newcommand{\inv}{$^{-1}$}
\newcommand{\kmps}{km s\inv}
\newcommand{\fdegr}{.\!\!\degr}

\newcommand{\cmcubic}{cm$^{-3}$}

\newcommand{\Jypb}{Jy beam\inv}
\newcommand{\mJypb}{m\Jypb}

\newcommand{\IcontInf}{66\degr} 

\newcommand{\Vsys}{3.9 \kmps}
\newcommand{\MdiskCO}{2.0 \Msun}
\newcommand{\IdiskCO}{70\degr}
\newcommand{\PAdiskCO}{50\degr}

\newcommand{\RoutdiskCO}{300 au} 
\newcommand{\RindiskCO}{50 au}

\newcommand{\MdiskCOInf}{1.5 \Msun}
\newcommand{\MdiskCOSup}{2.5 \Msun} 
\newcommand{\IdiskCOInffromModel}{40\degr}
\newcommand{\IdiskCOInf}{\IcontInf} 
\newcommand{\IdiskCOSup}{90\degr}

\newcommand{\VCOblueInf}{$-3$ \kmps}
\newcommand{\VCOblueSup}{$+3$ \kmps}

\newcommand{\Mire}{1.0 \Msun}
\newcommand{\Iire}{80\degr}
\newcommand{\CBire}{50 au}
\newcommand{\PAire}{50\degr}

\newcommand{\RoutireCO}{300 au}

\newcommand{\MireInf}{0.6 \Msun}
\newcommand{\MireSup}{1.2 \Msun}
\newcommand{\IireInf}{70\degr}
\newcommand{\IireSup}{90\degr}

\newcommand{\Vsysdisk}{2.5 \kmps}
\newcommand{\Mdisk}{0.4 \Msun}
\newcommand{\Idisk}{60\degr}
\newcommand{\PAdisk}{50\degr}
\newcommand{\PAdiskperp}{140\degr}
\newcommand{\Routdisk}{30 au}
\newcommand{\Rindisk}{0.56 au (1 pixel)}

\newcommand{\MdiskInf}{0.4 \Msun}
\newcommand{\MdiskSup}{0.8 \Msun}
\newcommand{\IdiskInf}{40\degr}
\newcommand{\IdiskSup}{70\degr}

\newcommand{\MireTFA}{0.20 \Msun}
\newcommand{\IireTFA}{70\degr}
\newcommand{\RoutireTFA}{30 au}
\newcommand{\CBireTFA}{10 au}

\newcommand{\CBireTFAtestInf}{5 au}
\newcommand{\CBireTFAtestSup}{20 au}

\newcommand{\MireTFAInf}{0.15 \Msun}
\newcommand{\MireTFASup}{0.25 \Msun}
\newcommand{\IireTFAInf}{60\degr}
\newcommand{\IireTFASup}{90\degr}

\newcommand{\VTFAInf}{$-6$ \kmps}
\newcommand{\VTFASup}{$+11$ \kmps}

\newcommand{\ltsim}{\protect\raisebox{-0.5ex}{$\:\stackrel{\textstyle <}{\sim}\:$}}

\newcommand{\remContinuum}{Contours represent the 1.3 mm continuum map. 
					For the contour levels, see the caption for Figure \ref{fig:mom0}.} 
\newcommand{\remChan}{The center velocity for each panel is shown in the upper-left corner, 
					where the systemic velocity of Source A is \Vsys.} 
\newcommand{\remContour}[3]{Contour levels 
					are at intervals of {#1}$\sigma$ starting from {#2}$\sigma$, 
					where the rms noise level is {#3} \mJypb.} 
\newcommand{\remContourPV}[3]{Contour levels 
					{\bff of the position-velocity diagrams} 
					are at intervals of {#1}$\sigma$ starting from {#2}$\sigma$, 
					where the rms noise level is {#3} \mJypb.}

\newcounter{tbnotecount}
\shorttitle{IRAS 16293-2422 A Substructures}
\shortauthors{Oya et al.}

\title{Substructures in the Disk-Forming Region of the Class 0 Low-Mass Protostellar Source IRAS 16293$-$2422 Source A on a 10 au Scale}

\author{
Yoko Oya\altaffilmark{1, 2} 
and Satoshi Yamamoto\altaffilmark{1, 2}
}

\email{oya@taurus.phys.s.u-tokyo.ac.jp}
\altaffiltext{1}{Department of Physics, The University of Tokyo, 7-3-1, Hongo, Bunkyo-ku, Tokyo 113-0033, Japan}
\altaffiltext{2}{Research Center for the Early Universe, The University of Tokyo, 7-3-1, Hongo, Bunkyo-ku, Tokyo 113-0033, Japan}

\begin{abstract}
We have observed the Class 0 protostellar source IRAS 16293$-$2422 A 
in the C$^{17}$O and H$_2$CS lines as well as the 1.3 mm dust continuum 
with the Atacama Large Millimeter/submillimeter Array 
at an angular resolution of $\sim$0\farcs1 (14 au). 
The continuum emission of the binary component, Source A, 
reveals the substructure consisting of 5 intensity peaks within 100 au from the protostar. 
The C$^{17}$O emission mainly traces
the circummultiple structure 
on a 300 au scale 
centered at the intensity centroid of the continuum, 
while it is very weak within the radius of 50 au from the centroid. 
The H$_2$CS emission, in contrast, traces the rotating disk structure around one of the continuum peaks (A1). 
Thus, it seems that the rotation centroid of the 
circummultiple structure 
is slightly different from that of the disk around A1. 
We derive the rotation temperature 
by using the multiple lines of H$_2$CS. 
As approaching to the protostar A1, 
the rotation temperature steeply rises up to 300 K or higher 
at the radius of 50 au from the protostar. 
It is likely due to a local accretion shock and/or the preferential protostellar heating 
of the transition zone from the 
circummultiple structure 
to the disk around A1. 
This position corresponds to the place 
where the organic molecular lines are reported to be enhanced. 
Since the rise of the rotation temperature of H$_2$CS most likely represents the rise of the gas and dust temperatures, 
it would be related to the 
chemical characteristics of this prototypical hot corino.

\end{abstract}


\begin{document}
\section{Introduction} \label{sec:intro}

Recently, rotationally-supported disks are found not only in Class I sources 
but also in some Class 0 sources 
\citep[e.g.][]{Yen2013, Yen2017, Murillo2013, Ohashi2014, Tobin_Perseus, Tobin2016a, Tobin2016b, Seifried2016, Lee2017_HH212, Aso2017, Okoda2018}. 
In spite of these {\bff extensive} efforts, 
it is still controversial when and how a disk structure is formed around a newly born protostar. 
Moreover, the disk formation process {\bff has been revealed} 
to be much more complicated for binary and multiple cases 
{\bff both in observations 
{\bff \citep{Tokuda2014, Dutrey2014, Takakuwa2014, Takakuwa2017, Tobin2016a, Tobin2016b, Boehler2017, ArturdelaVillarmois2018, Alves2019}} 
and numerical simulations 
\citep[e.g.][]{Bate1997, Kratter2008, Fateeva2011, Shi2012, Ragusa2017, Satsuka2017, Price2018, Matsumoto2019}.} 
For instance, 
circumbinary/circummultiple disk structures with a spiral structure 
as well as a circumstellar disk for each component are reported 
{\bff \citep[e.g.][]{Tobin2016b, Takakuwa2017, ArturdelaVillarmois2018, Matsumoto2019, Alves2019}.} 
In addition, it is not clear how molecules are processed during the disk formation process 
and what kinds of molecules are finally inherited to protoplanetary disks and potentially to planets. 
Understanding these processes is crucial, 
as they will provide important constraints on the initial physical and chemical conditions for the planetary-system formation study. 
In this context, 
{\bff {\bff physical} and chemical structures and their {\bff mutual} relation for} 
\dfr s of low-mass protostellar sources have been investigated 
with ALMA (Atacama Large Millimeter/submillimeter Array) 
\citep[e.g.][]{Sakai_1527nature, Sakai_1527apjl, Oya_16293, Oya_483, Oya_16293B, Oya_Elias29, Imai2016, Imai2019, Jacobsen2019}. 
These studies reveal 
that infalling envelopes and rotationally-supported disks 
are not smoothly connected to each other 
both in physical structure and chemical composition unlike previous expectations.

\iras\ is a well-studied low-mass protostellar source located in Ophiuchus 
\citep[$d \sim$ 140 pc;][]{Ortiz-Leon_Ophiuchus}. 
It is a Class 0 binary system, consisting of Source A and Source B, 
{\bff which are separated by 5\arcsec.} 
Both the components are famous for their richness in complex organic molecules (COMs) 
and are known as hot corinos 
\citep[e.g.][]{vanDishoeck1995, Cazaux2003, Schoier_hotcore, Bottinelli2004, Kuan2004, Ceccarelli2004, Chandler2005, Caux2011, Jorgensen_sugar}. 
Recently, the chemical characteristics of Source B has extensively been investigated 
by the PILS (Protostellar Interferometric Line Survey) program with ALMA 
{\bff \citep[e.g.][]{Coutens2016, Jorgensen_PILS, Lykke2017, Manigand2020b}.} 
{\bff As well, the characteristics of Source A has recently been reported by \citet{Manigand2020a} in the same program.} 
Meanwhile, 
Source A was investigated in terms of the kinematic structures; 
the rotating signature of the gas surrounding the protostar was detected thanks to its nearly edge-on configuration 
\citep[e.g.][]{Pineda_ALMA, Favre_SMA}. 
\citet{Oya_16293} reported that the kinematic structure of the gas in the vicinity of the protostar in Source A 
can be disentangled into the two major components; the \ire\ and the rotating disk. 
They evaluated the protostellar mass ($M = 0.75$ \Msun) 
and the specific angular momentum of the \ire\ ($j = 1.3 \times 10^{-3}$ \kmps\ pc) of this source 
under the assumption of the \ia\ of 60\degr\ (\incRem). 
The transition from the \ire\ to the rotating (Keplerian like) disk was found to be occurring 
at the radius of $(40-60)$ au from the center of gravity, 
which corresponds to the radius of the \cb\ of the \ire. 
The chemical composition of the gas was found to change drastically 
{\bff in} this transition zone.

The chemical evolution in protostellar sources is tightly related to the physical condition, 
and especially, the gas temperature distribution is of particular importance for its understanding. 
The temperature distribution in \iras\ 
has recently been modeled by \citet{Jacobsen2018_16293temp}; 
their 3D model of the dust and gas surrounding \iras\ Source A and Source B 
explains the observed distributions of the CO isotopologue lines and the dust emission. 
The gas dynamics connecting Source A and Source B has also been reported by \citet{vanderWiel2019}.
{\bff As for Source A, 
\citet{Oya_16293} and \citet{vantHoff2020} derived the temperature structure on a $(100 - 200)$ au scale 
by analyzing the intensity ratios of $K$-structure lines of \TFA\ observed at a 0\farcs5 resolution ($\sim$70 au). 
The result is discussed in relation to the drastic chemical change found on a 50 au scale mentioned above \citep{Oya_16293}.} 
Detailing a possible substructure in the transition zone is 
important for understanding 
the initial condition of the physical/chemical evolution in the disk formation process. 
However, the previous observations \citep{Oya_16293, vantHoff2020} 
do not have a sufficient spatial resolution ($\sim$0\farcs5; 70 au) 
{\bff for this purpose.} 
In this study, 
we make use of the ALMA data observed at a high spatial resolution ($\sim$0\farcs1; 14 au) 
and have a close look {\bff at} the disk/envelope structure of \iras\ Source A 
with particular emphasis on its substructure and gas-temperature distribution. 
{\bff We note that 
\citet{Maureira2020} have very recently reported a high resolution observation at 3 mm with ALMA, 
which can be compared with this study.} 

{\bff We summarize the observation in Section \ref{sec:obs} 
and its overall results in Section \ref{sec:results}. 
We present the kinematic structure of the gas around Source A in Section \ref{sec:discussion}, 
and the gas temperature distribution in Section \ref{sec:disc_temp}. 
Then, we discuss possible interpretations for the \desys\ of Source A in Section \ref{sec:disc_transition}. 
The major conclusions are summarized in Section \ref{sec:summary}.}

\section{Observations} \label{sec:obs}
Our observations toward \iras\ were carried out on 
21st August 2017  
with ALMA during its Cycle 4 operation. 
Forty-four antennas were used in the {\bff observations.} 
We employed the Band 6 receiver to observe 
the spectral lines of \CO\ and \TFA\ (Table \ref{tb:lines}). 
The field center of the observations was set to \iras\ Source A, 
which is located at $(\alpha_{\rm ICRS}, \delta_{\rm ICRS}) = (16^{\rm h}32^{\rm m}22\fs8713, -24\degr28\arcmin36\farcs502)$. 
The baseline lengths of the antennas ranged 
from 21.0 to 3696.9 m. 
The size of the field of view was 25\farcs15, 
and the typical size of the synthesized beam was $\sim$0\farcs1 for each image (see Table \ref{tb:lines}). 
The largest recoverable angular scale was 
2\farcs27. 
The total on-source time was 
70.6 minutes. 
Four spectral windows were observed. 
Their spectral resolution and the band width are summarized in Table \ref{tb:tuning}. 
The bandpass calibration was performed with J1517-2422, 
while the phase calibration was performed with J1625-2527 every 2 minutes. 
J1517-2422 and J1733-1304 were observed to derive the absolute flux density scale. 
The absolute accuracy of the flux calibration {\bff is} expected to be {\bff better} than 15 \%\ \citep{ALMA_TH-C4}. 

The continuum and line images were obtained with the CLEAN algorithm. 
We employed the Briggs weighting with a robustness parameter of 0.5, 
{\bff unless otherwise noted.} 
The 1.3 mm continuum image was prepared by averaging line-free channels, 
{\bff whose total frequency range was 0.3 GHz.} 
The line maps were obtained after subtracting the continuum component 
directly from the visibility data. 
The line maps were resampled to make the channel width to be 0.2 \kmps. 
A primary beam correction was applied to the continuum and the line maps. 
The root-mean-square (rms) noise level is 0.3 \mJypb\ for the continuum map, 
while it is 2.5 and 2.0 \mJypb\ for the \CO\ and \TFA\ line images, respectively. 
The velocity channel maps of the molecular lines were obtained by smoothing 
the velocity width of 1 \kmps\ so that the rms noise level is 1.1 and 0.9 \mJypb\ for the \CO\ and \TFA\ line images, resepctively. 
{\bff We tried self-calibration in various ways. 
However, the images are not improved or even deteriorated. 
Since we discuss the proper motion of the continuum peaks, 
we employ the images without self-calibration in this paper.} 

\section{Results} \label{sec:results}

\subsection{1.3 mm Continuum} \label{sec:results_cont}

Figure \ref{fig:cont}(a) shows the 1.3 mm dust continuum image of \iras. 
The continuum emission of Source A extends along the northeast-southwest direction (Figure \ref{fig:cont}b). 
Its elongated shape suggests the nearly edge-on configuration of Source A 
\citep[e.g.][]{Huang2005}. 
Moreover, it shows substructures. 
We {\bff identify} five intensity peaks, 
as shown in Figure \ref{fig:cont}(b). 
Their positions and {\bff peak} intensities are listed in Table \ref{tb:cont_peaks}. 
These values are obtained by using the 2D Fit Tool of {\tt casa viewer}. 
{\bff Although \iras\ is known to have a bridge structure connecting between Source A and Source B 
\citep[e.g.][]{Jorgensen_PILS, vanderWiel2019}, 
such a structure is missing in our observation probably due to the resolving-out effect.} 

{\bff 
In order to have a careful look at the clumpy structure of the 1.3 mm continuum emission, 
we prepare the continuum image cleaned with a uniform weighting {\bff for a better spatial resolution.} 
{\bff Figure \ref{fig:cont}(c)} depicts the continuum image 
obtained with the uniform weighting. 
The angular resolution of the image with a uniform weighting is 
(0\farcs106 $\times$ 0\farcs064) ($\sim$ 14 au $\times$ 9 au), 
{\bff which is slightly better than that with the {\bff Briggs} weighting
(0\farcs128 $\times$ 0\farcs080; $\sim$ 18 au $\times$ 11 au).} 
With the uniform weighting, 
the intensity peaks A1, A2, A3, and A4 are clearly resolved as well as in {\bff Figure \ref{fig:cont}(b).} 
{\bff In addition,} 
the peak A1a 
spatially separated from the nearby peak A1 
{\bff is tentatively detected.} 
{\bff Figure \ref{fig:cont}(d)} shows the spatial profile of the continuum emission 
along the line passing the peak A1 and A1a. 
The separation between A1 and A1a is clearer in the image with the uniform weighting. 
Although
the uniform weighting {\bff (Figure \ref{fig:cont}c)} provides a slightly better resolution 
than the Briggs weighting {\bff (Figure \ref{fig:cont}b),} 
the resolved-out effect increases due to less contribution of short-baseline data. 
Because we are interested in the structure surrounding Source A, 
we hereafter employ the continuum image with the Briggs weighting {\bff (Figures \ref{fig:cont}b),} 
which better traces the extended \desys. 
}

{\bff The two continuum peaks in Source A (A1, A2) 
were reported in the 2 cm (15 GHz) continuum image 
observed with VLA (Very Large Array) by \citet{Wootten1989}.}
{\bff More} 
clumpy {\bff features of Source A} 
has recently been reported in the 2 cm (15 GHz) and 0.9 cm (33 GHz) 
continuum images observed with VLA by \citet{HernandezGomez2019_contVLA}. 
{\bff Figures \ref{fig:cont_cm}(a) and (b)} show their cm continuum images overlaid on our mm continuum image. 
{\bff 
They reported the proper motion of these continuum peaks 
based on the VLA data performed from 1986 to 2015 
\citep[][]{Chandler2005, Pech2010},} 
except for the peak A2$\delta$. 
{\bff Figure \ref{fig:cont_cm}(c)} shows the extrapolation of the proper motion 
from the VLA observation in February 2014 (2014.15) to our ALMA observation in August 2017 (2017.64). 
Although calibration errors in the observations may affect the peak positions, 
the continuum peaks A1 and A2 observed with VLA in 2014 seem to correspond to those observed with ALMA in 2017. 
{\bff Very recently, 
the proper motion of A1 and A2 is also discussed by \citet{Maureira2020}.} 

It has been thought that 
{\bff Source A is a candidate of a multiple system 
based on its multiple outflow structures 
{\bff \citep[e.g.][]{Stark2004, Yeh2008, vanderWiel2019, Maureira2020}.} 
In fact, the continuum peaks A1 and A2 have been 
interpreted as protostars constituting a close binary system 
\citep{Loinard2007, Pech2010, HernandezGomez2019_contVLA}, 
although A1 was also suggested to be a shock feature 
\citep{Wootten1989, Chandler2005}.} 
On the other hand, 
the continuum peaks A2$\alpha$ and A2$\beta$ observed with VLA 
are interpreted as bipolar ejecta moving away from the protostar A2 
\citep[e.g.][]{HernandezGomez2019_contVLA}. 
{\bff These} two components (A2$\alpha$ and A2$\beta$) seem missing in our ALMA observation 
considering {\bff their} proper motion, 
{\bff although the continuum peak A3 detected with ALMA coincides with 
the A2$\beta$ position observed in 2014 \citep{HernandezGomez2019_contVLA}.} 
The continuum peaks A3 and A4 (and A1a) detected with ALMA 
do not seem to have corresponding components in the VLA images. 
{\bff 
These peaks are not clearly seen in the 3 mm continuum image by \citet{Maureira2020}, 
although a weak extended emission is seen around them. 
On the other hand, 
the peak A3 seems to correspond to Ab 
in the 1.3 mm continuum image reported 
by \citet{Sadavoy2018} 
\citep[see also][]{Chen2013, Chandler2005}.} 
The nature of the continuum peaks will be discussed later 
(Sections \ref{sec:disc_cont} and \ref{sec:disc_kin_tfa}).

\subsection{Molecular Lines} \label{sec:results_line}

In this observation, we detected various molecular lines 
as in the other observations toward this source \citep[e.g.][]{Jorgensen_PILS}. 
Among them, we focus on the \CO\ and \TFA\ lines in this study to characterize the gas distribution and the temperature distribution. 
Their line parameters are summarized in Table \ref{tb:lines}. 
Figure \ref{fig:mom0} shows their integrated intensity maps.

\subsubsection{C$^{17}$O} \label{sec:results_co}

The distribution of the \CO\ (\co) line is extended along the distribution of the continuum emission 
over 300 au in diameter (Figure \ref{fig:mom0}a). 
Although the signal-to-noise ratio of the integrated intensity map is not so high partly due to the resolved-out effect, 
the \CO\ emission seems to trace the 
{\bff the outer region of} the continuum distribution. 
Figure \ref{fig:mom0}(b) shows the velocity map (moment 1 map) of the \CO\ line. 
The velocity gradient is clearly seen along the northeast-southwest direction; 
the emission is blue-shifted in the northeastern side, 
while it is red-shifted in the southwestern side. 
The velocity gradient is almost parallel to the elongation of the distribution of the continuum emission. 
It seems to trace the rotating motion, 
which is consistent with the previous works
\citep[e.g.][]{Pineda_ALMA, Favre_SMA, Oya_16293}. 

Figure \ref{fig:chan_co} shows the velocity channel maps of the \CO\ line. 
The \CO\ emission is detected with the confidence level of 10$\sigma$ (11 \mJypb) or higher 
for the velocity range from $-1$ to 9 \kmps, 
or the velocity shift between $\pm5$ \kmps\ from the \sysV\ of \Vsys. 
Additional emission features are seen 
in panels for the velocity range from $-8$ to $-2$ \kmps\ and from 13 to 15 \kmps. 
These are contaminations of 
the (CH$_3$)$_2$CO ($18_{5, 13}$--$17_{6, 12}$ EE and $18_{6, 13}$--$17_{5, 12}$ EE at 224.700 GHz) transitions, 
the CH$_2$DOH ($7_{2, 6}$--$7_{1, 7}$, e$_1$ at 224.701 GHz; $20_{1, 19}$--$20_{0, 20}$, e$_1$ at 224.725 GHz) lines, 
{\bff and/or} an unidentified line. 
An absorption feature of the \CO\ line is seen around the \sysV, 
which is likely due to the self-absorption by the cold foreground gas. 
As seen in the velocity map (Figure \ref{fig:mom0}b), 
the rotation signature can be confirmed {\bff in the velocity channel maps.}

\subsubsection{H$_2$CS} \label{sec:results_h2cs}

Figures \ref{fig:mom0}(c), (d), and  (f) show the distributions of the \TFA\ lines. 
We detected the six $K$-structure lines of \TFA\ with the upper energy levels from 46 to 375 K as listed in Table \ref{tb:lines}, 
although the highest $K$ transitions (\tfaf) are 
contaminated with the \tfaa\ line 
(see Section \ref{sec:disc_kin_tfa}). 
All the \TFA\ line emissions show their intensity peak near the A1 continuum peak 
in contrast to the \CO\ (\co) line. 
Figure \ref{fig:mom0}(e) shows the velocity map of the \TFA\ (\tfab) line. 
The velocity gradient due to the rotation is clearly seen as the \CO\ line case. 

Figure \ref{fig:chan_h2cs} shows the velocity channel maps of the \TFA\ (\tfab) line. 
Here, we employ this line 
because the low-excitation line (\tfaa) is contaminated with the \TFA\ (\tfaf) line 
(see Section \ref{sec:disc_kin_tfa}). 
In contrast to the \CO\ (\co) line case, 
a self-absorption feature is not seen near the \sysV. 
The \TFA\ (\tfab) emission is detected with the confidence level of 10$\sigma$ (9 \mJypb) or higher 
for the velocity range from $-5$ to $+14$ \kmps, 
or the velocity shift {\bff from $-9$ to} $+10$ \kmps. 
The velocity range is wider than that of the \CO\ (\co) line. 
{\bff The high velocity blue-shifted emission is present on the northeastern side of A1, 
while the high velocity red-shifted emission is on the southwestern side. 
They likely traces the rotating motion around the protostar A1.} 
On the other hand, 
a clump is seen between the continuum peaks A2 and A4 
in panels for the velocity from 8 to 12 \kmps. 
This component can also be seen in the integrated intensity maps of the \TFA\ lines 
as a shape of `tongue' from A1 (Figures \ref{fig:mom0}c, d, f). 
{\bff It is highly red-shifted near the continuum peak A4 
in the velocity map (Figure \ref{fig:mom0}e).} 
We discuss this component in {\bff Sections \ref{sec:disc_kin_tfa_add}} and \ref{sec:disc_temp}.



\section{Analysis and Discussion} \label{sec:discussion}

\subsection{Substructures of the 1.3 mm Continuum Emission} \label{sec:disc_cont}

{\bff 
As {\bff described} in Section \ref{sec:results_cont}, 
\iras\ Source A is a 
multiple system 
including at least two protostars A1 and A2 
{\bff (Figure \ref{fig:cont}).} 
Observational researches of binary and multiple systems have recently been reported 
\citep{Tokuda2014, Takakuwa2014, Takakuwa2017, Tobin2016a, Tobin2016b, Boehler2017, ArturdelaVillarmois2018, Alves2019}. 
The radio continuum emission observed toward binary/multiple systems 
shows various substructures; 
for instance, 
L1551 NE has a circumstellar disk for each constituent of the binary 
and arm structures surrounding the binary system \citep{Takakuwa2017}, 
while [BHB2007] 11 has circumstellar disks connected to a circumbinary disk by spiral streamers \citep{Alves2019}. 
Numerical simulation studies for circumstellar disks indeed reproduce such substructures 
\citep[e.g.][]{Bate1997, 
Fateeva2011, 
Satsuka2017, Price2018, Matsumoto2019}. 
Both the observational and theoretical researches suggest a spatial gap {\bff of the gas distribution}  
between the circumbinary disk and the circumstellar disks. 
These structures are thought to be caused by gravitational instability 
due to rotation and/or turbulence of a parent core 
\citep[e.g.][]{Bate2003, Matsumoto2003, Lim2016}. 

In {\bff Figure \ref{fig:cont_cm},} 
the 1.3 mm continuum emission traces substructures similar to those observed previously with VLA 
{\bff \citep{HernandezGomez2019_contVLA}.} 
The {\bff weak extended} component on a 300 au scale shown in 
{\bff Figure \ref{fig:cont}(b)}  
seems to be a circumbinary/circummultiple envelope or disk, 
while each protostar inside it may be associated with a circumstellar disk revealed by a local emission peak. 
Our current observation does not show an {\bff apparent} spiral or arm structure potentially existing 
in the circummultiple structure, 
although the continuum peak A1a extended from A1 may be a hint of such inner structures. 
In the following sections, 
we investigate the physical and kinematic structures of 
the circummultiple structure 
and possible circumstellar disks. 
}

\subsection{Morphological Structure of the Disk/Envelope System} \label{sec:disc_kin}

The overall distribution of the 1.3 mm dust continuum emission in Source A looks like an ellipse (Figure \ref{fig:cont}), 
whose major axis extends along the northeast-southwest direction. 
It is necessary to define the center position and the major and minor axes of the distribution 
in order to investigate the kinematic structure of the gas {\bff showing} the rotation signature. 
However, this process is not straightforward, 
because the dust continuum distribution has substructures. 
Thus, we derive them in the following way. 
First, we determine the centroid of the continuum intensity distribution 
using the data points with 3$\sigma$ (0.9 \mJypb) detection or higher 
{\bff to be the following:} 
$(\alpha_{\rm ICRS}, \delta_{\rm ICRS}) = (16^{\rm h}32^{\rm m}22\fs873, -24\degr28\arcmin36\farcs614)$. 
The data points used for this calculation are within the magenta contour in Figure \ref{fig:centroid}(a). 
Then, we fit the positions of the data points used above to an ellipse by using the least-squares method 
weighted by the intensity at each position. 
The centroid of the ellipse is found to coincide with the intensity centroid obtained above within the pixel size of the map (0\farcs025). 
The major and minor axes are derived in the above fit to be 
1\farcs25 ($\sim$170 au) and 0\farcs50 ($\sim$70 au), respectively, 
while the position angle (P.A.) of the major axis of the elliptic distribution to be 
$50\fdegr2$. 
{\bff This ellipse is shown in {\bff Figure \ref{fig:centroid}(a).}} 
Thus, we assume that the mid-plane of the \desys\ extends along the P.A. of \PAire\ (hereafter `the envelope mid-plane direction'). 
{\bff If a flat disk structure without thickness is assumed,} 
its \ia\ is roughly estimated to be \IcontInf\ (\incRem) 
from the ratio of the sizes of the major and minor axes. 
This value can be regarded as a lower limit to the \ia, 
if a finite thickness {\bff and a round shape} for the disk structure {\bff are} considered. 
Thus, the \desys\ of \iras\ Source A is confirmed to be nearly or almost edge-on as reported previously 
\citep[e.g.][]{Pineda_ALMA, Favre_SMA, Oya_16293}. 
The envelope mid-plane direction and the intensity centroid determined above are 
shown in Figure \ref{fig:centroid}(a) by a white arrow and a black cross, respectively. 
The intensity centroid can be regarded as a rough estimate of the center of gravity for the elliptic structure, 
assuming that the intensity of the dust emission is proportional to the dust mass at each position.



Figure \ref{fig:spatialProfile}(a) shows the spatial profiles 
of the continuum intensity and the integrated intensities of the molecular lines 
along the envelope mid-plane direction. 
The continuum emission shows a peak at the angular offset of 0\arcsec\ (the intensity centroid of the continuum emission), 
which corresponds to the skirt of the A1 peak. 
It also has an additional peak at the angular offset of $+0\farcs3$ ($\sim$40 au), 
which corresponds to that of the A3 peak.  
Similarly, the \TFA\ (\tfaa; \tfab; \tfad) emissions have 
a peak at the angular offset of 0\arcsec\ with a shoulder or another peak at the offset of 0\farcs2. 
Thus, these emissions are centrally concentrated. 
On the other hand, 
the \CO\ (\co) emission is more extended than the continuum and the \TFA\ emissions 
and shows a double-peaked feature 
with a depression of its intensity in the area inward of 50 au from the centroid 
(the interior of the two dotted lines in Figure \ref{fig:spatialProfile}a), 
where the continuum and the \TFA\ emissions are the brightest.

\subsection{Kinematic Structure of the \CO\ Line} \label{sec:disc_kin_co}

We first focus on the double-peaked distribution of the \CO\ (\co) emission 
and analyze its kinematic structure. 
{\bff Figure \ref{fig:PV_co-Kepler}} shows the position-velocity (PV) diagrams of the \CO\ line 
prepared along the 6 directions shown by arrows in the {\bff attached} velocity map, 
which are centered at the intensity centroid of the continuum emission.
In the PV diagram along the envelope mid-plane direction {\bff (Figure \ref{fig:PV_co-Kepler}a),} 
we see the spin-up feature toward the intensity centroid of the continuum emission; 
{\bff the velocity gets more red-shifted from the \sysV\ (\Vsys) 
as approaching to the protostar position from the southwestern side, 
while it gets more blue-shifted as approaching from the northeastern side.} 
Interestingly, this feature {\bff almost} vanishes inward of 0\farcs4 ($\sim$50 au) from the continuum peak position. 
{\bff This corresponds to the position where the integrated intensity of the \CO\ line decreases (Section \ref{sec:disc_kin}; {\bff Figure \ref{fig:spatialProfile}a).}} 

{\bff 
We also inspected the velocity structure of the \CO\ line around the continuum peak A1 
as in the case of the \TFA\ lines, 
which will be described in Section \ref{sec:disc_kin_tfa}. 
We found that 
the \CO\ emission does not show a symmetric feature around A1 in contrast to Figure \ref{fig:PV_co-Kepler}. 
Therefore, 
we concluded that the center for the kinematic structure of the \CO\ emission 
should be 
at the center of gravity of the elliptic structure 
traced by the dust continuum emission, 
rather than at A1. 
} 

{\bff In this section, we compare the observed kinematic structure with two simple models which have often been used in previous studies.} 
{\bff We here use three-dimensional physical models 
of a flat disk/envelope with the Keplerian motion or the infalling-rotating motion. 
The details for these models are reported by \citet{Oya_15398} \citep[][]{Oya_PhD, Oya_FERIA}.
In these models, 
the line emission is assumed to be optically thin, 
and the radiation transfer is not considered. 
Free parameters are the protostellar mass and the \ia\ for the Keplerian model, 
while the \ire\ model has the radius of the \cb\ as a free parameter in addition to them.} 
We conduct the chi-squared ($\chi^2$) test for {\bff the two models} 
to obtain the reasonable parameters. 
{\bff The details for the reduced $\chi^2$ test are described in \ref{sec:app_chisq}.} 
{\bff The best fit and the reasonable ranges for the parameters 
are summarized in Table \ref{tb:modelparams}.} 

\subsubsection{Keplerian Model for the \CO\ Line} \label{sec:disc_kin_co_Kep}

First, 
we examine the Keplerian model for the PV diagrams of \CO. 
We {\bff optimize} 
{\bff the protostellar mass and the \ia\ as free parameters} 
to explain the observed PV diagrams. 
%
{\bff Although the reduced $\chi^2$ values suffer from 
the imperfection of the model and the complexity of the sources,}
we obtain the {\bff following} best fit parameters of the Keplerian model: 
the central mass is \MdiskCO\ and the \ia\ is \IdiskCO, 
{\bff as shown in Table \ref{tb:modelparams}.} 
{\bff Here, the inner and outer radii of the model are fixed to be \RindiskCO\ and \RoutdiskCO\ 
according to the observed molecular distribution.} 
{\bff 
Other details 
of the fittings are described in \ref{sec:app_chisq_co}.} 

Figure \ref{fig:PV_co-Kepler} shows the PV diagrams of the Keplerian model with the best fit parameters. 
As depicted in Figure \ref{fig:PV_co-Kepler}(a), 
the PV diagram along the envelope mid-plane direction is reasonably reproduced. 
However, the PV diagrams along the other directions are not well explained, 
although their overall trend is {\bff roughly} reproduced (Figures \ref{fig:PV_co-Kepler}b--f). 
{\bff The resolved-out effect and/or the self-absorption effect would affect 
the velocity components near the \sysV\ (\Vsys); 
the observation does not show emission while the model does. 
On the other hand, 
there are some parts in the PV diagrams 
where the line emission is detected but is not reproduced by the model. 
Such discrepancy originates from the imperfection of the model.} 
For instance, the model cannot well explain the PV diagram 
along the direction perpendicular to the envelope mid-plane direction (Figure \ref{fig:PV_co-Kepler}d). 
The blue-shifted component in Figure \ref{fig:PV_co-Kepler}(d) has a higher velocity-shift 
in the observation than the model.

\subsubsection{Infalling-Rotating Envelope Model for the \CO\ Line} \label{sec:disc_kin_co_IRE}

{\bff 
Next, 
we examine the \ire\ model \citep{Oya_15398}. 
In this model, the velocity field is approximated by the ballistic motion. 
The energy and specific angular momentum of the gas is assumed to be conserved, 
and thus the gas cannot fall inward of the periastron, 
which is called as the `\cb'. 
%
The {\bff following} best fit parameters for the \ire\ model {\bff are obtained}: 
the protostellar mass is \Mire\ and the \ia\ is \Iire. 
{\bff The radius of the \cb\ and the outer radius of the model are fixed to be \CBire\ and \RoutireCO\ 
according to the observed molecular distribution.} 
{\bff The results for the fittings are described in \ref{sec:app_chisq_co}.} 

{\bff Figure \ref{fig:PV_co-IRE} shows the PV diagrams of the \ire\ model with the best fit parameters.} 
{\bff As in the case of the Keplerian disk model described above, 
the \ire\ model does not explain the observed kinematic structure very well, either;}
{\bff for instance, 
the velocity-shift in the model is larger than the observed kinematic structure 
in Figures \ref{fig:PV_co-IRE}(c), (d), and (e).} 
The PV diagrams along the envelope mid-plane direction is {\bff reasonably} reproduced. 
However, 
the red-shifted component of the \CO\ (\co) line seems to severely suffer from the absorption effect by the infalling gas, 
so that a velocity gradient due to the infall motion 
expected along the direction perpendicular to the envelope mid-plane direction 
in the model is not evident in Figure \ref{fig:PV_co-IRE}(d). 

\subsubsection{Circummultiple Structure} \label{sec:disc_kin_co_circummultiple}

{\bff 
Neither of the two simple models fully explains the detailed kinematic structure because of the complexity of the source. 
This {\bff source has} the small substructures, 
which are now resolved in our observation. 
Hence, the kinetic structure would be too complicated to be modeled by these simple models. 
{\bff Although the \ire\ model gives a smaller $\chi^2$ values than the Keplerian model (Tables \ref{tb:chi2_C17O-Kep} and \ref{tb:chi2_C17O-IRE}), 
it would be too hasty to conclude that 
the \CO\ emission comes from an \ire.} 
{\bff Nevertheless, 
we can roughly constrain the central mass of Source A 
from the kinematic structure observed at the 0\farcs1 resolution. 
Finer tuning of the model by, 
for instance, combining the two simple models 
{\bff is practically difficult, 
because such an analysis includes too many free parameters for our current observational data of this source, 
and it} is out of scope of this work.}

Then, we compare the protostellar masses {\bff derived} above with those in the literatures. 
The protostellar mass of Source A is reported to be from 0.5 \Msun\ to 1.0 \Msun\ 
\citep[e.g.][]{Looney2000, Bottinelli2004, Huang2005, Caux2011, Pineda_ALMA, Favre_SMA, Oya_16293}, 
while \citet{Takakuwa2007b} {\bff and \citet{Maureira2020}} reported a larger central mass from 0.5 \Msun\ to 2.0 \Msun. 
The central mass employed for the \ire\ model above (\Mire) is consistent with most of the previous reports. 
{\bff Meanwhile, the mass} 
employed for the Keplerian model (\MdiskCO) 
{\bff agrees with the highest end of the previous reports.} 

As mentioned in Section \ref{sec:intro}, 
\citet{Oya_16293} reported the \ire\ structure with a radius of 300 au around Source A in the OCS emission 
{\bff at an angular resolution of $(0\farcs65 \times 0\farcs51)$.} 
{\bff They suggested that 
the gas outside the radius of 50 au from the center of gravity is an \ire\ while 
that inside the radius of 50 au is the Keplerian disk. 
If the \CO\ gas has the Keplerian motion rather than the infalling-rotating motion, 
the \CO\ gas would have a different kinematic structure from the OCS gas 
at the radius from 50 au to \RoutdiskCO\ from the continuum intensity centroid,} 
{\bff although} the angular resolution of 
{\bff the OCS observation} 
is coarser than that of 
{\bff the \CO} 
observation. 
{\bff Thus,} 
the OCS distribution could be {\bff somewhat} different from the \CO\ distribution. 
{\bff For instance, if}  
the envelope gas falls onto a thin rotating disk diagonally from above and below the mid-plane 
\citep[e.g. Figures 4.5 and 4.11 in][]{Hartmann2009}, 
the OCS (\ocs) line with the high upper-state energy (111 K) 
will {\bff preferentially} trace a warm surface of the infalling envelope gas. 
In contrast, 
the \CO\ (\co) line with the upper-state energy of 16 K 
tends to trace a cold dense mid-plane of the rotating disk, 
{\bff whose motion is close to be Keplerian.} 

The \CO\ emission seems to trace 
the circummultiple structure in Source A 
and is deficient in the vicinity of the center of gravity. 
A circumbinary disk is previously observed for L1448-IRS3B 
by \citet{Tobin2016b} in the {\bff $^{13}$CO} (\co) emission with ALMA. 
Meanwhile, 
a hole of the {\bff $^{13}$CO distribution} tracing the rotating gas 
is previously reported for TMC-1A 
by \citet{Harsono2018}. 
{\bff 
They interpreted the hole structure as absorption of the $^{13}$CO emission by 
the optically thick dust.} 
{\bff On the other hand,} 
theoretical simulations show  
a gap of {\bff the gas distribution} between a circumbinary disk and circumstellar disks 
\citep[e.g.][]{Bate1997, 
Fateeva2011, 
Satsuka2017, Price2018, Matsumoto2019}. 
{\bff At the first glance,} 
such a gap/hole structure would not be the case of the \CO\ emission in \iras\ Source A; 
{\bff in the \CO\ deficient area, 
the dust 
is not expected to be so optically thick as to significantly suppress the molecular emission enough,} 
because the \TFA\ emission is observed {\bff there (Figure \ref{fig:mom0}).} 
{\bff However, an optically thick dust may cause the depression of the \CO\ emission 
if the molecular distribution is different between CO and \TFA\ (see next section for details).} 


\subsubsection{\CO\ Deficiency within 50 au} \label{sec:disc_kin_co_def}

The weak \CO\ (\co) emission near the centroid of the continuum emission 
(Figures \ref{fig:mom0}a and \ref{fig:spatialProfile}a) 
is {\bff difficult to be} 
interpreted by the 
{\bff freezing out} of CO {\bff onto dust grains.} 
In this area, 
the gas temperature is expected to be much higher than the desorption temperature of CO (20 K) 
according to the analysis of the \TFA\ line (see Section \ref{sec:disc_temp}). 
If the dust temperature is also as high as the gas temperature, 
the weak emission of the \CO\ (\co) line cannot simply be attributed to the adsorption of CO onto dust grains.

The adsorption of CO might be the case, 
{\bff if} the disk mid-plane temperature 
{\bff within \CBire} 
were lower than 
{\bff the adsorption temperature of CO.} 
On the other hand, 
the \TFA\ emission would mainly come from the warm region including the disk surface, 
and the effect of the depletion in the disk mid-plane could be less important. 
This picture qualitatively explains 
the behavior of the spatial profiles shown in Figure \ref{fig:spatialProfile}(a).

{\bff 
If such different distributions between CO and \TFA\ are the case, 
the optical depth effect of the continuum emission affects the \CO\ emission more seriously 
than the \TFA\ emissions. 
The \CO\ emission from the cold and dense mid-plane 
would be depressed by an optically thick dust, 
while the \TFA\ emissions from the warm surface region would be less affected. 
Such an effect of the optically thick dust was invoked 
for the {\bff $^{13}$CO} 
hole observed in TMC-1A by \citet{Harsono2018}. 
}

{\bff 
The excitation effect can cause the weak \CO\ emission in the hot region in the vicinity of the continuum peak A1. 
In Figure \ref{fig:spatialProfile}(b), 
the integrated intensity is 49 \mJypb\ \kmps\ (120 K \kmps) and less than 3$\sigma$ (21 \mJypb\ \kmps; 51 K \kmps) 
at its peak and A1, respectively. 
If the column density is the same for these two positions, 
the difference between these integrated intensities is explained 
by a factor of three difference of the excitation temperature 
(i.e., 100 K and 300 K for these two positions). 
Here the local thermodynamical equilibrium (LTE) condition and the optically thin condition are assumed. 
In this case, 
the excitation temperature would change sharply as the distance from the protostar, 
since the intensity of the \CO\ emission sharply changes. 
}

Alternatively, 
non-volatile organic molecules formed in the gas-phase of the disk component are adsorbed onto dust grains, 
which may result in the exhaustion of carbon from the gas-phase \citep{Aikawa1996}. 
This might eventually cause deficiency of CO. 
{\bff In fact, \iras\ Source A is known to be rich in COMs. 
Moreover, \citet{Oya_16293} reported that the COM emission is enhanced within the radius of \CBire, 
where the CO emission is weak in this observation.} 
The deficiency in CO 
may be related to such peculiar chemical characteristics.

\subsection{Kinematic Structure of the \TFA\ Lines} \label{sec:disc_kin_tfa}
\subsubsection{Rotating Disk around A1} \label{sec:disc_kin_tfa_A1}

In contrast to the \CO\ line, 
the \TFA\ lines are concentrated 
within the radius of \CBire\ (Figures \ref{fig:mom0}c, d, f). 
In order to investigate this feature, 
we prepare the integrated intensity maps of the \TFA\ (\tfab) line 
for the blue- and red-shifted velocity ranges {\bff (Figure \ref{fig:centroid}b).} 
Note that this line {\bff is} free from {\bff significant} contamination lines {\bff by other lines} 
according to the spectral line database \citep[JPL and CDMS;][]{Pickett_JPL, Muller_CDMS, Endres_CDMS}, 
which is also confirmed in the actual observations 
{\bff by \citet{vantHoff2020} and this work.} 
Figure \ref{fig:centroid}(b) shows the high velocity-shift components with the velocity shift from $\pm4$ to $\pm8$ \kmps. 
The rotation motion can be recognized around the continuum peak A1 
rather than the intensity centroid of the dust continuum, 
although an additional red-shifted component is seen near the continuum peak A4. 
The blue-shifted and red-shifted components overlap just at the continuum peak A1. 
This result indicates the rotating structure associated with A1. 
Since the \desys\ of Source A shows substructures in its continuum emission, 
the center of gravity could be different between the envelope and this rotating structure.  
By focusing on the vicinity of A1, 
the P.A. of the velocity gradient due to the rotation is evaluated to be \PAdisk. 
Thus, we assume that the mid-plane of the rotating structure around A1 extends along this P.A., 
which is parallel to the envelope mid-plane direction derived from the dust continuum distribution (Section \ref{sec:disc_kin}). 
This direction (hereafter `the disk mid-plane direction') is shown by a red arrow in Figure \ref{fig:centroid}(b). 
Figure \ref{fig:spatialProfile}(b) shows the spatial profiles 
of the continuum emission and the integrated intensities of the molecular lines 
along the disk mid-plane direction. 
The continuum emission and the \TFA\ emission are concentrated to the continuum peak A1, 
while the \CO\ emission shows a double-peaked feature, 
as {\bff the spatial profiles along the envelope {\bff mid-plane} direction} (Figure \ref{fig:spatialProfile}a).

\subsubsection{Keplerian Model for the \TFA\ Line} \label{sec:disc_kin_tfa_Kep}

{\bff 
First, we perform the $\chi^2$ test for the Keplerian model and the \TFA\ (\tfad) line, 
as the \CO\ case in Section \ref{sec:disc_kin_co_Kep} (see also \ref{sec:app_chisq}).} 
{\bff 
We obtain the {\bff following} best fit {\bff parameters}: 
the central mass is \Mdisk\ and the \ia\ is \Idisk, 
where 
{\bff the \sysV\ {\bff is} \Vsysdisk. 
The inner and outer radii of the model are fixed to be $\sim$0 au and \Routdisk\ 
according to the observed molecular distribution, respectively.} 
{\bff We regarded the emission extending over 30 au 
as a part of the circummultiple structure traced by the \CO\ emission (Section \ref{sec:disc_kin_co}), 
and we did not take them into account in the following analysis.} 
{\bff The best fit and the reasonable ranges for the parameters are summarized in Table \ref{tb:modelparams}.} 
{\bff Other details for the fittings are described in \ref{sec:app_chisq_tfa}.}

{\bff Figure \ref{fig:PV_h2cs-707-725-744-Kep} shows} 
the PV diagrams of the \TFA\ (\tfaa; \tfab; \tfad) lines 
{\bff along the directions with P.A.s of 50\degr\ and 80\degr\ centered at the protostar A1.} 
{\bff The} Keplerian model {\bff results with the best fit parameters obtained above} is overlaid 
{\bff on the observation results.} 
{\bff The PV diagrams of \TFA\ (\tfad) along all the direction shown by 
the black arrows 
in the velocity map (moment 1 map) on the top left panel are 
shown in 
\ref{sec:app_chisq_tfa} (see Figure \ref{fig:PV_h2cs-744-Kep}).} 
We see the high velocity-shift components concentrated to the protostar A1. 
The PV diagrams {\bff in Figure \ref{fig:PV_h2cs-707-725-744-Kep} as well as those in Figure \ref{fig:PV_h2cs-744-Kep}} 
can be {\bff reasonably} explained by the Keplerian model described above, 
except for the {\bff three} features particularly seen in {\bff Figures \ref{fig:PV_h2cs-707-725-744-Kep}(b) and (d)} 
{\bff (see Section \ref{sec:disc_kin_tfa_add}).} 
%

\subsubsection{Infalling-Rotating Envelope Model for the \TFA\ Line} \label{sec:disc_kin_tfa_IRE}

{\bff 
Next, we performe the $\chi^2$ test for the \ire\ model and the \TFA\ (\tfad) line, 
as the \CO\ case in Section \ref{sec:disc_kin_co_IRE}. 
We obtain the {\bff following} best fit parameters: 
the central mas is \MireTFA\ and the \ia\ is \IireTFA, 
where the \sysV\ is \Vsysdisk. 
The outer radius of the model is fixed to be \RoutireTFA\ 
according to the observed molecular distribution. 
The best fit and the reasonable ranges for the parameters are summarized in Table \ref{tb:modelparams}. 
Other details for the fittings are described in \ref{sec:app_chisq_tfa}.
As the result, we find that 
the kinematic structure traced by \TFA\ is worse reproduced by the \ire\ model 
than by the Keplerian model (Table \ref{tb:modelparams}; see Figure \ref{fig:PV_h2cs-744-IRE}). 
Thus, the observed kinematic structure would prefer the Keplerian motion. 
%
%
%
}

\subsubsection{Comments on Additional Features} \label{sec:disc_kin_tfa_add}

{\bff 
According to the model analysis in Sections \ref{sec:disc_kin_tfa_Kep} and \ref{sec:disc_kin_tfa_IRE}, 
it is most likely that the continuum peak A1 is a protostellar source 
associated {\bff with} a rotating disk structure. 
{\bff Some components detected in the \TFA\ emission 
are not attributed to this rotating disk structure. 
Figures \ref{fig:PV_h2cs-707-725-744-Kep}(a) and (c) show 
low velocity-shift components extended from the angular offset of $\pm$0\farcs2 to $\pm$1\arcsec. 
As we have described in Section \ref{sec:disc_kin_tfa_Kep}, 
this component can be attributed to the circummultiple structure traced by the \CO\ emission (Figures \ref{fig:PV_co-Kepler}, \ref{fig:PV_co-IRE}).} 
{\bff Moreover,} 
the observed PV diagrams show three features spilling over the model results, 
as we mentioned in Section \ref{sec:disc_kin_tfa_Kep}. 
} 

{\bff An excess emission is seen} 
at the velocity of about 12 \kmps\ toward the A1 position (zero offset) 
{\bff in the PV diagram of the \TFA\ (\tfaa) line {\bff (`Excess 1' in Figure \ref{fig:PV_h2cs-707-725-744-Kep}a, b).}} 
Since this feature is not seen in {\bff the \TFA\ (\tfad) (Figure \ref{fig:PV_h2cs-707-725-744-Kep}e, f)} 
line, 
this emission is most likely interpreted as contamination by the unresolved doublet of \TFA\ (\tfaf). 
Although \citet{Oya_16293} suggested an existence of the Keplerian disk component 
based on the observation of the \TFA\ (\tfaa) line, 
it would {\bff have suffered} from this contamination \citep[see also][]{vantHoff2020}. 
{\bff 
The \TFA\ (\tfab) line shows a weak emission (`Excess 2') near Excess 1, too. 
Since it has a slight offset from the A1 position, 
it may be a contamination by another molecular line 
rather than a high velocity-shift component of the \TFA\ (\tfab) line in the vicinity of the protostar. 
The peak intensity of this weak contamination is less than 30 \%\ of that of the \TFA\ (\tfab) line, 
and moreover, it is separated from the \TFA\ (\tfab) line along the velocity axis. 
Therefore,} 
the \TFA\ (\tfab) line is verified to have no significant contamination by other molecular lines 
in its PV diagrams {\bff (Figure \ref{fig:PV_h2cs-707-725-744-Kep}c, d),} 
{\bff as noted in the previous sections.} 
{\bff As well, 
this weak contamination does not affect the discussion 
for the rotation temperature of the \TFA\ in Section \ref{sec:disc_temp}.} 
Thus, we can certainly confirm the disk component with this line.} 
It should be noted that 
the detection of the high-excitation line of \TFA\ (a composite of \tfaf; \Eu\ $=$ 375 K) suggests a high gas temperature 
of the disk component near the protostar A1. 

{\bff Another excess emission} 
{\bff in Figure \ref{fig:PV_h2cs-707-725-744-Kep}(b)} is the high-velocity component at an offset of $-0\farcs5$ from A1 
{\bff (`Excess 3').} 
This component can be {\bff seen} in {\bff the \TFA\ (\tfab, \tfad)} 
lines {\bff (Figure \ref{fig:PV_h2cs-707-725-744-Kep}d, f),} 
and 
{\bff cannot be ascribed to the contamination of the other lines.} 
This component {\bff is} 
neither the \ire\ described in Section \ref{sec:disc_kin_co} nor the disk component around A1. 
It likely represents the component extended through the midway between A2 and A4 
(see Section \ref{sec:results_h2cs}). 
This feature is interesting, 
because the velocity increases as an increasing distance from the protostar A1 
({\bff Figure \ref{fig:mom0}e;} see the $v =$ 7--{\bff 12} \kmps\ panels of {\bff Figure \ref{fig:chan_h2cs}).} 
It might be a part of the outflowing motion. 
Alternatively, it looks like a part of a spiral structure (Figure \ref{fig:mom0}f), 
as seen in circumbinary systems 
\citep{Takakuwa2014, Takakuwa2017, Matsumoto2019, Alves2019}. 
Further characterization of this component is left for future study. 

\subsubsection{Comparison between the Circumstellar and Circummultiple Structures} \label{sec:disc_kin_tfa_compare}

{\bff 
We have also examined a possible disk structure associated with the other continuum peaks. 
For instance, 
Figure \ref{fig:PV_h2cs-744_A2} shows the PV diagrams of the \TFA\ (\tfad) line, 
where the position axes are centered at the protostar A2. 
If a molecular line traces a possible disk associated to A2, 
the PV diagrams have to be symmetric with respect to the A2 position and 
the \sysV\ of A2 (See Figures {\bff \ref{fig:PV_co-Kepler} and} \ref{fig:PV_co-IRE} for reference). 
However, such an expected feature is not {\bff evident} in Figure \ref{fig:PV_h2cs-744_A2}. 
One may think that 
the kinematic structure in the panel (a) (P.A. of \PAdisk) 
{\bff showing a velocity gradient} 
can be interpreted as the Keplerian motion 
with a \sysV\ of $\sim$6 \kmps. 
{\bff In fact, 
\citet{Maureira2020} has recently suggested the existence of the rotating disk structure associated with the protostar A2.} 
{\bff However,} 
it is more adequately interpreted as a part of other components, 
i.e. the disk associated to A1 or the circummultiple structure 
of Source A 
{\bff in our observational result.} 
{\bff At the current stage, 
we cannot conclude whether a rotating disk structure exists around the protostar A2.} 
{\bff As well as the protostar A2, 
we have investigated the kinematic structure in the vicinity of the continuum peaks A3 and A4.  
Our current observation did not show a rotating disk structure around either of them.}

{\bff Although the physical parameters in Table \ref{tb:modelparams} 
are obtained with the simplified models, 
it is worth noting that the best fit 
{\bff masses, \ia s, and systemic velocities} 
are different 
between the circummultiple structure of Source A traced by the \CO\ (\co) emission 
and the circumstellar structure of the protostar A1 traced by the \TFA\ (\tfad) emission. 
First, the central mass estimated for the circummultiple structure is 
from \MdiskCOInf\ to \MdiskCOSup\ with the Keplerian disk model 
and from \MireInf\ to \MireSup\ with the \ire\ model. 
These values are larger than the central mass 
from \MdiskInf\ to \MdiskSup\ {\bff (the Keplerian model) and from \MireTFAInf\ to \MireTFASup\ (the \ire\ model)} 
estimated 
for the circumstellar structure {\bff around A1.} 
The circummultiple structure surrounds 
possible other {\bff components} (A2, A3, A4, and A1a) as well as the protostar A1, 
and {\bff also} their possible circumstellar structures. 
{\bff Thus, the central mass evaluated from} the kinematic structure of the circummultiple structure 
{\bff should be higher than the mass evaluated for A1.} 

{\bff 
\citet{Maureira2020} have recently reported that 
the gas mass is $(1-3) \times 10^{-3}$ \Msun, $(1-3) \times 10^{-3}$ \Msun, and $(0.03-0.1)$ \Msun\ 
for the substructure around A1, that around A2, and the extended structure surrounding both of them, respectively, 
based on the 3 mm continuum emission. 
These values are significantly smaller than the difference of the central mass 
between the circummultiple and circumstellar structures found in this study. 
Thus, the mass of the rest sources, 
i.e. the protostar A2 and possible protostars A3 and A4, 
would contribute to the kinematic structure of the circummultiple structure.} 

Second, the \ia\ seems larger for the circummultiple structure than the circumstellar structure, 
{\bff although it has a} large uncertainty.
A1 is likely one component of a potential binary/multiple system of Source A. 
Thus, the {\bff differences} of the \ia\ and the \sysV\ imply 
that the rotating disk structure around A1 
could be tilted with respect to the circummultiple structure 
of Source A. 
As well, the difference of the \sysV\ between the two models, 
\Vsys\ for the circummultiple structure and \Vsysdisk\ for the circumstellar structure, 
implies a potential motion of the protostar A1 in Source A. 
}

\section{Distribution of the Rotation Temperature of \TFA} \label{sec:disc_temp}

{\bff In this section, 
we investigate the temperature distribution using the high spatial resolution data of \TFA. 
So far, the temperature structure on a $(100 - 200)$ au scale have been studied by \citet{Oya_16293} and \citet{vantHoff2020} 
by using the $K$-structure lines of \TFA\ observed at a 70 au resolution. 
The rotation temperature derived from these lines is a good measure of the gas kinetic temperature, 
because the radiation processes between the different $K_{\rm a}$ levels ($\Delta K_{\rm a} = \pm2, \pm4$) are very slow. 
Hence, they can be used for delineating the temperature structure around the protostar. 
\citet{Oya_16293} derived the rotation temperature at the five positions 
with offsets of $\pm 1\arcsec$, $\pm 0\farcs5$, and 0\arcsec\ {\bff from the continuum peak position,} 
which {\bff correspond} to the envelope, the transition zone (\cb), and the disk, respectively, 
by using the intensity of the \tfaa, \tfab, and \tfada/\tfadb\ lines 
integrated over the velocity range corresponding to each component. 
Based on the result, 
they suggested the temperature rise at the transition zone by comparison of the temperatures at the five positions 
and {\bff discussed}  
the result in relation to the enhancement of COMs near the transition zone. 
Later, \citet{vantHoff2020} observed more \TFA\ lines to derive the temperature structure of this source. 
They use the peak fluxes instead of the integrated intensity 
to derive the line ratio and delineate the temperature distribution in Source A. 
The temperature profile along the direction of the \desys\ does not reveal the local temperature rise 
suggested by \citet{Oya_16293}. 
Thus, they {\bff interpreted} the temperature profile in terms of the radiation heating from the protostar. 
Although these works reveal the fundamental temperature distribution of this source, 
the spatial resolution is too coarse to examine the temperature structure around the transition zone in detail. 
Since substructures in the circummultiple system of Source A are now evident, 
we revisit the temperature structure of this source with the high resolution \TFA\ data. 
We here employ the \TFA\ (\tfab) and (\tfad) lines for the analysis, 
because} the \TFA\ (\tfaa) line suffers from the contamination by the \TFA\ (\tfaf) lines. 
{\bff As we describe in Section \ref{sec:disc_kin_tfa_add}, 
the weak contamination for the \TFA\ (\tfab) line (Excess 2 in Figure \ref{fig:PV_h2cs-707-725-744-Kep}c, d) 
does not matter 
as long as we discuss the line intensities with PV diagrams or velocity channel maps.} 
The unresolved doublet of \TFA\ (\tfad) is confirmed to have no significant contamination 
{\bff in the molecular line database and this observation (Figure \ref{fig:PV_h2cs-744_A2}),} 
as the \TFA\ (\tfab) line {\bff case} mentioned in Section {\bff \ref{sec:disc_kin_tfa_add}.}

Figure \ref{fig:mom0_h2cs-Trot}(a) shows the distribution of the ratio of the integrated intensity 
of the \TFA\ (\tfad) line relative to that of the \TFA\ (\tfab) line, 
{\bff while} 
Figure \ref{fig:mom0_h2cs-Trot}(b) shows the distribution of the rotation temperature of \TFA\ calculated 
from the integrated intensity ratio. 
{\bff Figures \ref{fig:spatialProfile_intensity}(a) and \ref{fig:spatialProfile_Trot}(a) show 
{\bff the} spatial profiles {\bff of the integrated intensity ratio and the rotation temperature} 
along the disk mid-plane direction (a red arrow in Figure \ref{fig:centroid}b), {\bff respectively.}} 
Here, 
we assume {\bff the LTE condition and} 
the optically thin conditions for the two lines. 
{\bff The LTE condition for \TFA\ is well fulfilled in this hot and dense region, 
as revealed with the non-LTE calculation by \citet{Oya_16293}.} 
{\bff In order to verify the optically thin condition, 
we roughly calculate the optical depths of the \TFA\ (\tfab) and (\tfad) lines 
from the observed brightness temperature and the derived rotation temperature. 
{\bff Here, we ignore the contribution of the dust emission, 
because the treatment of this effect requires detailed radiation transfer 
and is out of the scope of this paper. 
The derived optical depths} are mostly lower than 0.5, 
and its median is 0.28 and 0.10 
for the \TFA\ (\tfab) and (\tfad) lines, respectively. 
The optical depth of the \TFA\ (\tfab) line exceeds 1 for some parts. 
However, such high optical depth parts are in the low temperature regions ($<$100 K). 
Since we discuss the temperature structure of hot components around the protostar, 
the optical depth effect does not affect seriously.} 

The rotation temperature {\bff of \TFA\ derived} 
{\bff from the \tfab\ and \tfada/\tfadb\ lines} 
{\bff is} confirmed to be consistent 
with {\bff the gas kinetic temperature} derived by using RADEX code \citep{vanderTak_radex} 
with the H$_2$ density range of $10^{7-9}$ \cmcubic, 
which is relevant to \desys s \citep{Oya_16293}. 
The rotation temperature of \TFA\ is {\bff evaluated to be} as high as 300 K near A1. 
{\bff A high rotation temperature area is extended from A1 to southwestern part. 
The 
rotation temperature apparently 
decreases as the increasing distance from the 
{\bff protostar} A1 position 
{\bff with a slight enhancement at the southwest part.} 
{\bff These features are} consistent with the result obtained at a lower spatial resolution by \citet{vantHoff2020}.} 
It is worth noting that the tongue-like feature from A1 (Section \ref{sec:results_h2cs}) 
shows high temperature ($>$200 K; Figure \ref{fig:mom0_h2cs-Trot}b). 
As discussed in Section \ref{sec:disc_kin_tfa}, 
this might be related to the outflowing motion.


{\bff 
We have also derived the rotation temperature from the intensity ratio of the \tfab\ line to the \tfaa\ line for reference. 
The temperature is generally lower than that derived from the \tfab\ and \tfada/\tfadb\ lines. 
For instance, 
the former is 100 K and the latter is 200 K at the continuum peak A3. 
This is probably because the \tfaa\ line would sample colder regions 
and is partly contaminated by the \TFA\ (\tfaf) line (Section \ref{sec:disc_kin_tfa_add}). 
Hence, we use the rotation temperature derived from the \tfab\ and \tfada/\tfadb\ lines 
in the following discussions. 
}

Figure \ref{fig:chan_h2cs-Trot} shows the velocity channel maps of the rotation temperature of \TFA. 
Each map is derived based on the intensity ratio of the velocity channel maps of the \TFA\ (\tfab) and (\tfad) lines. 
As in Figure \ref{fig:mom0_h2cs-Trot}(b), 
the optically thin conditions for the two lines and the LTE condition are assumed. 
In Figure \ref{fig:mom0_h2cs-Trot}, 
the envelope and disk components are contaminated with each other along the \los, 
because the line emission is integrated for the velocity axis. 
On the other hand, 
they are expected to be disentangled to some extent in the velocity channel maps. 
In Figure \ref{fig:chan_h2cs-Trot}, 
the rotation temperature of \TFA\ apparently 
{\bff becomes higher as approaching to} 
the {\bff protostar} A1 position. 
{\bff It is as high as} 
400 K near A1. 

In the panels for the velocity of $-3$ and $+11$ \kmps\ of Figure \ref{fig:chan_h2cs-Trot}, 
the rotation temperature is as high as $\sim$400 K 
at some positions apart from the protostar A1. 
{\bff These positions roughly correspond to the positions 
within which the \CO\ emission decreases. 
Figures \ref{fig:mom8_h2cs-Trot}(a) and (b) show the maps of 
the {\bff highest} intensity ratio of the two \TFA\ lines and 
the {\bff highest} rotation temperature 
along the velocity channels, 
which are derived from the cube data. 
In other words, the maximum values along the \los\ are shown in these maps. 
{\bff These maps are obtained by using the {\tt immoments} task of {\tt casa} 
with the option ``{\tt moments = [8]}''.} 
{\bff Figure \ref{fig:spatialProfile_intensity}(b) shows their spatial profiles 
along the disk mid-plane direction.} 
The maximum intensity ratio  {\bff along the \los} 
steeply increases 
at the distance of 50 au from the protostar A1, 
and are almost constant within this radius. 
The highest rotation temperature along the \los\ shown in {\bff Figure \ref{fig:spatialProfile_Trot}(b)} also steeply increases 
at the same position but shows a large scatter within it. 
{\bff It should be noted that} very high temperatures obtained at some positions {\bff would be} artificial. 
Since the intensity ratio at these positions are close to the high-temperature limit (1.47 for $T=\infty$) 
{\bff as shown in Figure \ref{fig:spatialProfile_intensity}(b),} 
the derived temperature is sensitive to a small change in the ratio 
and suffers from a large error. 
Therefore, the intensity ratio itself better represents the temperature profile. 
{\bff Interestingly, the positions of the steep increase corresponds 
the transition zone from the circummultiple structure to the circumstellar structure, 
as mentioned {\bff above.} 


Thus, the local rise of the rotation temperature seems occurring near the transition zone 
from the circummultiple structure of Source A to the circumstellar structure of the protostar A1. 
A hint for such a temperature rise in the transition zone was previously suggested by \citet{Oya_16293}. 
{\bff However, this previous {\bff report would have been}  just 
{\bff fortuitous based on the results for a few positions,} 
given the detailed analysis {\bff by} \citet{vantHoff2020}.} 
Now, our {\bff high-resolution} observation ($\sim$0\farcs1) {\bff indeed reveals} a sharp change in the temperature.}  
The local rise of the rotation temperature of \TFA\ most likely reflects a local rise of the gas temperature. 
Candidate mechanisms causing a local temperature rise 
are discussed in Section \ref{sec:disc_transition}.

\section{Transition from the Envelope to the Disk-Forming Region} \label{sec:disc_transition}

When the radiation heating by the central protostar {\bff (A1)} is considered as a heating source, 
the gas temperature is expected to gradually decrease 
as a power-law of the distance to the protostar. 
In the edge-on case, the maximum gas temperature along the \los\ is 
expected to be as well (Figure \ref{fig:scheme_temp}a). 
{\bff If the protostars other than A1 could contribute to the heating, 
the temperature profile would be more gentle than the case of the heating only by A1.} 
{\bff This effect may contribute to flatten the temperature profile around A1 within the radius of 50 au. 
Nevertheless, 
it seems difficult to explain 
the local steep rise of the temperature on both the northeastern and southwestern sides of A1 
by the protostellar heating alone.} 
Thus, another heating mechanism 
is expected in \iras\ Source A. 

{\bff 
The peak temperature profile projected onto the plane of the sky is expected to be flat, 
if the {\bff actual} radial temperature profile is flat as well 
or it has a local rise in a ring-like structure.
The observed temperature profile, 
{\bff which shows a steep rise around 50 au and then flattens,} 
requires that 
the temperature is not significantly higher in the vicinity of the protostar than 
at the distance of 50 au from the protostar due to some mechanisms. 
The temperature is naturally expected to increase as approaching to the protostar by the radiation heating, 
and thus, a completely flat temperature profile within the radius of 50 au is unlikely. 
It seems more likely that 
the radial temperature profile has an increase at the radius of 50 au, 
drops just inside it, 
and gradually increases as approaching to the protostar. 
} 

If the local temperature rise just outside 
{\bff the circumstellar structure of A1} 
occurs due to an accretion shock by the infalling gas, 
it will appear {\bff in} a ring-like structure surrounding the disk structure (Figure \ref{fig:scheme_temp}b). 
{\bff In fact, 
the model study by \citet{Fateeva2011} shows 
that bow-shocks can occur {\bff near the inner edge of a circumbinary disk.} 
It may heat up the innermost edge of the circummultiple structure 
in \iras\ Source A.} 
In the edge-on or nearly edge-on case, 
the maximum gas temperature along the ling-of-sight 
is expected to show a flat feature as a function of the distance from the protostar (A1). 
{\bff This corresponds to what can be seen in Figures {\bff \ref{fig:spatialProfile_intensity}{\bff (b)} and} \ref{fig:spatialProfile_Trot}{\bff (b).}} 

Such a gas temperature distribution with a ring-like structure is also caused by a physical structure of the gas; 
the stagnated gas in the transition zone from 
{\bff the circummultiple structure to the curcumstellar structure} 
can be piled up vertically to the mid-plane of the disk, 
and the gas is efficiently heated up by the radiation from the protostar 
without the shielding by the {\bff circumstellar} disk structure (Figure \ref{fig:scheme_temp}b). 
In any case, 
the distribution of the rotation temperature in \iras\ Source A 
shown in Figure \ref{fig:mom8_h2cs-Trot}(b) 
seems consistent with {\bff the} picture {\bff of} the hot ring-like region. 

The picture of the hot ring-like region 
would also be related to the local enhancement 
of the COM emission reported by \citet{Oya_16293}. 
The \MN\ emission and the \MF\ emission likely come from 
{\bff the} ring-like {\bff region} with a radius of 50 au surrounding the protostar, 
based on their kinematic structures. 
According to \citet{Oya_16293}, 
the COM emission seems weaker in the {\bff circumstellar} disk structure than {\bff at the transition zone.} 
{\bff Although} the spatial-resolution {\bff of their observation} 
is insufficient, 
this distribution {\bff could} be caused by the temperature distribution shown in Figure \ref{fig:scheme_temp}(b). 
COMs may be re-depleted onto dust grains because of the lower temperature of the disk mid-plane. 
If this picture is the case, 
the `hot corino' would not have a simple spherical or disk-like structure in the vicinity of the protostar, 
but a ring-like structure surrounding the protostar. 
The gas and dust temperature will rise again in the closest vicinity of the protostar, 
and COMs can desorb from the dust grains there. 
{\bff Although one may think that the weak COM emission is due to the optically thick dust emission, 
it is not likely because the \TFA\ emission is strongly detected in the same area. 
As described in Section \ref{sec:disc_kin_co_def} for the \CO\ and \TFA\ case, 
this situation could be the case 
if \TFA\ and COM distribute in the gas phase of different parts of the \desys.} 
{\bff To} understanding {\bff the actual} situation, 
{\bff it is required 
to confirm whether the distribution of the COM line emission 
is really weak within a certain radius.} 
Observations of COM lines at a high spatial resolution 
will help us {\bff in this aspect,} 
which are in progress. 
{\bff Observations at a lower frequency are favorable 
for this purpose in order to avoid the effect of dust opacity.}

\section{Summary} \label{sec:summary}

We {\bff have} observed the Class 0 low-mass protostellar source \iras\ Source A 
in the \CO\ and \TFA\ lines as well as in the 1.3 mm dust continuum 
at a high angular-resolution of $\sim$0\farcs1 (14 au). 
The major findings are as follows: 

\begin{itemize}
\item[(1)] The 1.3 mm continuum emission traces the nearly edge-on \desys. 
		Moreover, the {\bff substructures} on a 0\farcs5 ($<$100 au) scale are resolved, 
		and 5 continuum peaks are identified. 
		The positions of two continuum peaks (A1 and A2) are consistent 
		with those previously observed at cm wavelength, 
		if their proper {\bff motions are} considered. 
		{\bff They also agree with the recent report on the 3 mm continuum emission.} 
		We also detect new continuum peaks (A3, A4, and A1a). 
		{\bff Meanwhile,} 
		A2$\alpha$ and A2$\beta$ observed at cm wavelength, 
		which are thought to be the ejecta from A2, 
		are missing. 
		Thus, \iras\ Source A is likely a multiple system. 
\item[(2)] The \CO\ (\co) line emission traces 
		{\bff the rotating gas} 
		on a 300 au scale 
		centered at the intensity centroid of the dust continuum emission. 
		On the other hand, 
		\CO\ {\bff is found to be} deficient 
		{\bff within the radius of \CBire\ from the intensity centroid of the continuum emission.} 
		{\bff Enhancement of the temperature, 
		depletion} of CO onto dust grains in the mid-plane of the disk/envelope structure, 
		and exhaustion of carbon to non-volatile organic molecules 
		{\bff can be considered as the origin of this feature.} 
\item[(3)] The multiple lines of \TFA\ {\bff mainly} trace 
		the disk component {\bff around the strongest continuum peak (A1).} 
		Even the high-excitation lines (\tfaf; \Eu\ $=$ 375 K) are detected. 
\item[(4)] The kinematic {\bff structures} of the \CO\ {\bff and \TFA\ lines are examined} 
		by 
		{\bff a Keplerian model and an \ire\ model.} 
		The center of gravity, the \ia, and the \sysV\ {\bff of the circummultiple structure {\bff of Source A} traced by \CO\ are} 
		{\bff possibly} 
		different {\bff from those of the 
		{\bff circumstellar structure} 
		around the {\bff protostar} A1 traced by \TFA.} 
\item[(5)] The distribution of the rotation temperature of \TFA\ is derived. 
		In an overall view, 
		the rotation temperature increases as approaching to the protostar A1, 
		as usually expected for the gas temperature profile determined by the radiation heating from the protostar. 
		{\bff Interestingly,} 
		the rotation temperature shows a steep rise around the transition zone 
		from {\bff the circummultiple structure} 
		to the {\bff circumstellar} disk {\bff associated to the protostar A1.}  
		This {\bff local} steep rise {\bff could} be attributed to a possible accretion shock by the infalling gas 
		and/or the thermal heating of the stagnated gas by the protostar. 
		If the dust temperature rises there as well, 
		it would be related to the rich COM emission, 
		which is the chemical characteristics of \iras\ Source A. 
		`Hot corino' may have a ring-like structure, 
		instead of a simple spherical or disk-like structure. 
\end{itemize}

While theoretical studies have extensively progressed in the disk-formation study 
\citep[e.g.][]{Hennebelle2009, 
Li2011, Machida2011, Tomida2015, Tsukamoto2017, Lam2019}, 
observational studies tend to have been behind them due to the limited high-resolution observations. 
In these years, the radio observational studies have revealed the fundamental picture of the \ire\ and the disk inside it, 
thanks to the unprecedented angular-resolution and the sensitivity of 
millimeter and submilimeter interferometers including ALMA. 
This work demonstrates that the transition zone from 
{\bff the circummultiple disk/envelope to the circumstellar disk is not smooth.} 
Moreover, 
even substructures of the \dfr\ are now in the scope of the observations. 
Further study on these issues will provide us with deep insight 
into the formation process of the disk and the chemical evolution during it.

\acknowledgements

{\bff The authors thank the anonymous referees for valuable suggestions to improve the paper.} 
The authors thank 
Drs. Nami Sakai, 
Ana L\'{o}pez-Sepulcre, 
Yoshimasa Watanabe, 
Cecilia Ceccarelli, 
Bertrand Lefloch, 
C\'{e}cile Favre, 
and 
Nienke van der Marel 
for their invaluable discussion 
in the initial stage of the project. 
This study used the ALMA data set ADS/JAO.ALMA\#2016.1.00457.S. 
ALMA is a partnership of the European Southern Observatory, 
the National Science Foundation (USA), 
the National Institutes of Natural Sciences (Japan), 
the National Research Council (Canada), 
and the NSC and ASIAA (Taiwan), 
in cooperation with the Republic of Chile. 
The Joint ALMA Observatory is operated by the ESO, the AUI/NRAO, and the NAOJ. 
The authors are grateful to the ALMA staff for their excellent support. 
This study is supported by a Grant-in-Aid from the Ministry of Education, 
Culture, Sports, Science, and Technologies of Japan 
(grant Nos. 18H05222, 19H05069, and 19K14753).

\appendix 

\renewcommand{\thesection}{Appendix \Alph{section}}

\section{Reduced Chi-Squared Test for the Position-Velocity Diagrams} \label{sec:app_chisq}

{\bff In Sections \ref{sec:disc_kin_co} and \ref{sec:disc_kin_tfa}, 
we conduct the chi-squared ($\chi^2$) test 
for the observed PV diagrams and the model results.} 
{\bff We calculate the {\bff reduced} $\chi^2$ value {\bff ($\chi^2$/DOF),} 
which is the sum of the square of the difference 
between the modeled and observed PV diagrams 
along the envelope mid-plane direction at each pixel 
{\bff normalized} by the square of the rms in the observation 
{\bff (2.5 \mJypb\ for \CO, 2.0 \mJypb\ for \TFA)}.} 

{\bff {\bff A} large reduced $\chi^2$ value {\bff mainly} originates 
from {\bff complexity of the source and} imperfection of the model. 
Hence, we should take relatively large ranges of the parameters 
which can reasonably reproduce the observed features. 
Based on this thought, 
the parameter ranges are determined from the range 
where the reduced $\chi^2$ value increases by 1. 
{\bff The best fit and the reasonable ranges for the parameters derived by the $\chi^2$ test 
are summarized in Table \ref{tb:modelparams}.}

\section{Analysis of \CO} \label{sec:app_chisq_co}

Since the \CO\ emission notably suffers from the self-absorption effect, 
we {\bff only use} the blue-shifted component from \VCOblueInf\ to \VCOblueSup\ to calculate $\chi^2$. 
%
Table \ref{tb:chi2_C17O-Kep} shows {\bff reduced} $\chi^2$ for the central-mass versus \ia\ plane. 
Figure \ref{fig:PV_co-Kepler_MI} shows 
examples of the PV diagrams of the Keplerian model overlaid on the \CO\ observation. 
We {\bff confirm} 
the \ia\ {\bff from \IdiskCOInffromModel\ to} \IdiskCOSup\ to be the reasonable range 
for 
the Keplerian model. 
We also consider the lower limit of the \ia\ of \IcontInf\ obtained from the distribution of the continuum emission (Section \ref{sec:disc_kin}), 
{\bff and thus the reasonable range for the \ia\ is 
from \IdiskCOInf\ to \IdiskCOSup.} 
{\bff Then, we confirm 
the central mass from \MdiskCOInf\ to \MdiskCOSup\ to be the reasonable range for the Keplerian model.} 
It should be noted that the protostellar mass and the \ia\ are correlated with each other, 
as shown by the boxes in Table \ref{tb:chi2_C17O-Kep}.

{\bff As the Keplerian model analysis described above,} 
we {\bff also investigate} 
{\bff the \ire\ model with the two variable} 
parameters 
(the central mass {\bff and the \ia)} 
to explain the observations 
{\bff by using the $\chi^2$ test. 
{\bff The radius of the \cb\ and the outer radius of the model 
are fixed to be \CBire\ and \RoutireCO\ based on the observed distribution.} 
Table \ref{tb:chi2_C17O-IRE} shows the results {\bff of the reduced $\chi^2$ test} 
{\bff for the central mass versus \ia\ plane.}  
{\bff Figure \ref{fig:PV_co-IRE_MI} shows 
examples of the PV diagrams of the \ire\ model overlaid on the \CO\ observation.} 
We {\bff confirm} 
the protostellar mass from \MireInf\ to \MireSup\ 
and the \ia\ from \IireInf\ to \IireSup\ to be the reasonable ranges 
for the parameters of the \ire\ model. 

\section{Analysis of \TFA} \label{sec:app_chisq_tfa}

In contrast to the \CO\ analysis, 
we use the two PV diagrams of the \TFA\ (\tfad) line 
prepared along the disk mid-plane direction (P.A. \PAdisk) and the direction perpendicular to it (P.A. \PAdiskperp), 
and use both the red- and blue-shifted components with the velocity range from \VTFAInf\ to \VTFASup. 

The results {\bff for the Keplerian model} are summarized in Table \ref{tb:chi2_h2cs744-Kep}. 
Figures \ref{fig:PV_h2cs-744-Kep_MI-PA050} and \ref{fig:PV_h2cs-744-Kep_MI-PA140} 
{\bff show} 
examples of the model results 
{\bff for the PV diagrams along and perpendicular to the disk mid-plane {\bff direction,} respectively.}} 
The reasonable ranges for the parameters are: 
the central mass is from \MdiskInf\ to \MdiskSup\ and 
the \ia\ is from \IdiskInf\ to \IdiskSup. 
{\bff We here assume 
the systemic velocity of \Vsysdisk\ 
and the outer radius of \Routdisk.} 
{\bff Figure \ref{fig:PV_h2cs-744-Kep} shows 
the model results with the best fit parameters 
overlaid on the observed PV diagrams of the \TFA\ line (\tfad)
prepared for various directions.}

{\bff The results for the \ire\ model are summarized in Table \ref{tb:chi2_h2cs744-IRE}.} 
Figures \ref{fig:PV_h2cs-744-IRE_MI-PA050} and \ref{fig:PV_h2cs-744-IRE_MI-PA140} 
show some examples of the model results 
for the PV diagrams along and perpendicular to the disk mid-plane {\bff direction,} respectively. 
The reasonable ranges for the paramters are: 
the central mass is from \MireTFAInf\ to \MireTFASup\ and 
the \ia\ is from \IireTFAInf\ to \IireTFASup. 
{\bff The radius of the \cb\ is varied from \CBireTFAtestInf\ to \CBireTFAtestSup. 
We here assume 
the systemic velocity of \Vsysdisk\ and the outer radius of \Routdisk.} 
{\bff Figure \ref{fig:PV_h2cs-744-IRE} shows 
the model results with the best fit parameters 
overlaid on the observed PV diagrams of the \TFA\ line (\tfad)
prepared for various directions.} 

The observed PV diagram along the disk mid-plane direction (P.A. 50\degr; Figure \ref{fig:PV_h2cs-744-IRE}a) 
seems reasonably reproduced by the \ire\ model, 
as in the case of the Keplerian model (Section \ref{sec:disc_kin_tfa_Kep}). 
On the other hand, 
some diagrams are worse reproduced by the \ire\ model than by the Keplerian model. 
For instance, 
the observational result in Figure \ref{fig:PV_h2cs-744-IRE}(b) shows a velocity gradient 
where 
the red- and blue-shifted components are 
on the southwestern and the northeastern sides of the protostar A1, respectively. 
The \ire\ model does not explain this observed trend well, 
while the Keplerian model does {\bff (Figures \ref{fig:PV_h2cs-707-725-744-Kep}f and \ref{fig:PV_h2cs-744-Kep}b).} 
{\bff In fact, 
the reduced $\chi^2$ value is larger for the fit by the \ire\ model 
than by the fit by the Keplerian model (Tables \ref{tb:chi2_h2cs744-Kep} and \ref{tb:chi2_h2cs744-IRE}).} 
Thus, the observed kinematic structure 
would prefer the Keplerian motion than the infalling-rotating motion.



\begin{landscape}
\begin{table}
	\begin{center}
	\caption{Observed Molecular Lines\tablenotemark{a}
			\label{tb:lines}}
	\begin{tabular}{lccccc}
	\hline 
	Molecule & Transition & Rest Frequency (GHz) & S$\mu^2$ ($D^2$) & \Eu\ (K) & Synthesized Beam Size \\ \hline \hline
	1.3 mm Continuum & & $222.45-240.58$ & & & $0\farcs128 \times 0\farcs080$ (P.A. $-83\fdegr729$) \\ 
	\CO & \co & 224.714385 & 0.014 & 16 & $0\farcs119 \times 0\farcs083$ (P.A. $89\fdegr205$) \\ 
	\TFA\tablenotemark{b} & \tfaa & 240.2668724 & 19.0 & 46 & $0\farcs118 \times 0\farcs077$ (P.A. $-85\fdegr842$) \\ 
	& \tfab & 240.5490662 & 17.5 & 99 & $0\farcs118 \times 0\farcs079$ (P.A. $-87\fdegr729$) \\ 
	& \tfac & 240.3820512 & 17.5 & 99 & $0\farcs118 \times 0\farcs078$ (P.A. $-85\fdegr873$) \\ 
	& \tfae & 240.3937618, 240.3930370 & 46.6, 46.6 & 165 & $0\farcs118 \times 0\farcs077$ (P.A. $-85\fdegr880$) \\ 
	& \tfad & 240.3321897 & 12.8, 12.8 & 257 & $0\farcs118 \times 0\farcs078$ (P.A. $-85\fdegr868$) \\ 
	& \tfaf & 240.2619875 & 28.0, 28.0 & 375 & $0\farcs118 \times 0\farcs077$ (P.A. $-85\fdegr842$) \\ 
	\hline 
	\end{tabular}
	\tablenotetext{a}{{\bff Rest frequency, S$\mu^2$, and \Eu\ for each line are} taken from CDMS \citep{Muller_CDMS, Endres_CDMS}.}
	\tablenotetext{b}{{\bff Original references: \citet{Maeda2008}, \citet{Muller2019}.}}
	\end{center}
\end{table}
\end{landscape}

\begin{table}
	\begin{center}
	\caption{Settings of the Spectral Windows in the Observation
			\label{tb:tuning}}
	\begin{tabular}{ccccc}
	\hline
	SPW ID & Frequency Range (GHz) & Resolution (kHz) & Molecular Lines \\ \hline \hline 
	0 & 240.2148020--240.4491915 & 61.039 & \TFA\ (\tfaa), (\tfac), \\ 
	 & & & (\tfae), \\ 
	 & & & (\tfad), \\
	 & & & and (\tfaf) \\  
	1 & 240.5167470--240.5753444 & 15.260 & \TFA\ (\tfab) \\ 
	2 & 224.6885482--224.8057430 & 30.519 & \CO\ (\co) \\ 
	3 & 222.4463399--223.3838980 & 488.311 & Continuum \\ 
	\hline 
	\end{tabular}
	\end{center}
\end{table}

\begin{table}
	\begin{center}
	\caption{Intensity Peaks in the 1.3 mm Continuum Map\tablenotemark{a}
			\label{tb:cont_peaks}}
	\begin{tabular}{lccc}
	\hline 
	Peak & Peak Intensity (mJy/beam) & RA (ICRS) & Dec (ICRS) \\ \hline \hline 
	A1 & 37.6 & 16$^{\rm h}$32$^{\rm m}$22\fs878 & $-$24\arcdeg28\arcmin36\farcs691 \\ 
	A1a & 28.5 & 16$^{\rm h}$32$^{\rm m}$22\fs873 & $-$24\arcdeg28\arcmin36\farcs785 \\ 
	A1a (Uniform weighting)\tablenotemark{b} & 19.2 & 16$^{\rm h}$32$^{\rm m}$22\fs873 & $-$24\arcdeg28\arcmin36\farcs827 \\ 
	A2 & 22.4 & 16$^{\rm h}$32$^{\rm m}$22\fs852 & $-$24\arcdeg28\arcmin36\farcs658 \\ 
	A3 & 24.0 & 16$^{\rm h}$32$^{\rm m}$22\fs887 & $-$24\arcdeg28\arcmin36\farcs496 \\ 
	A4 & 20.4 & 16$^{\rm h}$32$^{\rm m}$22\fs846 & $-$24\arcdeg28\arcmin36\farcs821 \\ 
	Intensity Centroid\tablenotemark{c} & & 16$^{\rm h}$32$^{\rm m}$22\fs873 & $-$24\arcdeg28\arcmin36\farcs614 \\
	\hline 
	\end{tabular}
	\tablenotetext{a}{Obtained from the continuum image with the Briggs weighting by using the {\tt 2D Fit Tool} of {\tt casa viewer}.} 
	\tablenotetext{b}{\bff Obtained from the continuum image with the uniform weighting. The beam size is 0\farcs106 $\times$ 0\farcs064 in this image.} 
	\tablenotetext{c}{Obtained as the averaged position for the 3$\sigma$ detection of the dust continuum emission in Source A. 
				See Section \ref{sec:disc_kin}.}
	\end{center}
\end{table}

\renewcommand{\baselinestretch}{1.2}

\begin{landscape}
\begin{table}
	\vspace*{-45pt}
	\begin{center}
	\caption{Physical Parameters Derived in the Model Analyses 
			\label{tb:modelparams}}
	\begin{tabular}{l|cccc}
	\hline
	& \multicolumn{2}{c}{\CO\ (\co)} & \multicolumn{2}{c}{\TFA\ (\tfad)} \\ 
	Parameters & Keplerian Model & IRE Model\tablenotemark{a} & Keplerian Model & IRE Model\tablenotemark{a} \\ \hline \hline 
	Center of Gravity\tablenotemark{b} & 
	\multicolumn{2}{c}{
	\begin{tabular}{c} 
	Intensity centroid \vspace*{-5pt} \\ 
	of the continuum emission 
	\end{tabular}
	}
	& 
	\multicolumn{2}{c}{Continuum peak A1} \\ 
	\hspace*{-10pt}
	\begin{tabular}{l}
	Coordinates (ICRS) \vspace*{-5pt} \\ of the Center of Gravity 
	\end{tabular}
	& 
	\multicolumn{2}{c}{
	\begin{tabular}{c}
	RA: $16^{\rm h}32^{\rm m}22\fs873$ \vspace*{-5pt} \\ Dec: $-24\arcdeg28\arcmin36\farcs614$ 
	\end{tabular}
	}
	& 
	\multicolumn{2}{c}{
	\begin{tabular}{c}
	RA: $16^{\rm h}32^{\rm m}22\fs878$ \vspace*{-5pt} \\ Dec: $-24\arcdeg28\arcmin36\farcs614$ 
	\end{tabular}} 
	\\ 
	Position Angle\tablenotemark{c} &\PAdiskCO & \PAire & \PAdisk & \PAdisk \\ \vspace*{-12pt} \\ 
	Systemic Velocity & \Vsys & \Vsys & \Vsysdisk & \Vsysdisk \\ \vspace*{-12pt} \\ 
	Inclination Angle\tablenotemark{d} & \IdiskCO & \Iire & \Idisk & \IireTFA \\ 
	\quad (Reasonable Range) & (\IdiskCOInf$-$\IdiskCOSup) & (\IireInf$-$\IireSup) & (\IdiskInf$-$\IdiskSup) & (\IireTFAInf$-$\IireTFASup) \\ \vspace*{-12pt} \\ 
	Central Mass\tablenotemark{e} & \MdiskCO & \Mire & \Mdisk & \MireTFA \\
	\quad (Reasonable Range) & (\MdiskCOInf$-$\MdiskCOSup) & (\MireInf$-$\MireSup) & (\MdiskInf$-$\MdiskSup) & (\MireTFAInf$-$\MireTFASup) \\ \vspace*{-12pt} \\ 
	Radius of the CB\tablenotemark{f} & $-$ & \CBire & $-$ & \CBireTFA \\ 
	Inner Radius & \RindiskCO & \CBire & \Rindisk & \CBireTFA \\ 
	Outer Radius\tablenotemark{g} 
	& \RoutdiskCO & 
	\RoutireCO 
	& \Routdisk 
	& \RoutireTFA \\ 
	Minimum $\chi^2$/DOF & 3.38 & 2.18 & 7.07 & 8.98 \\ 
	Figures & \ref{fig:PV_co-Kepler}, \ref{fig:PV_co-Kepler_MI} & \ref{fig:PV_co-IRE}, \ref{fig:PV_co-IRE_MI}
	& \ref{fig:PV_h2cs-707-725-744-Kep}, \ref{fig:PV_h2cs-744-Kep_MI-PA050}, \ref{fig:PV_h2cs-744-Kep_MI-PA140}, \ref{fig:PV_h2cs-744-Kep} 
	& \ref{fig:PV_h2cs-744-IRE_MI-PA050}, \ref{fig:PV_h2cs-744-IRE_MI-PA140}, \ref{fig:PV_h2cs-744-IRE} 
	\\ \hline 
	\end{tabular}
	\vspace*{-25pt}
	\tablenotetext{a}{Infalling-rotating envelope model.} 
	\tablenotetext{b}{See Figure \ref{fig:centroid}.}
	\tablenotetext{c}{The position angle of the direction along which the \desys\ extends.}
	\tablenotetext{d}{\incRem.
					\bff The values in the parentheses show the reasonable range of the \ia\ obtained by the {\bff reduced $\chi^2$} test 
					(Tables \ref{tb:chi2_C17O-Kep}--\ref{tb:chi2_h2cs744-IRE}) 
					{\bff and the distribution of the continuum emission (see Section \ref{sec:disc_kin})}.}
	\tablenotetext{e}{Note that the central mass and the \ia\ are correlated with each other: 
					the central mass is approximately scaled by ${\displaystyle \frac{1}{\sin^2 i}}$, where $i$ is the \ia\ (\incRem).
					\bff The values in the parentheses show the reasonable range of the central mass obtained by the {\bff reduced $\chi^2$} test 
					(Tables \ref{tb:chi2_C17O-Kep}--\ref{tb:chi2_h2cs744-IRE}).}
	\tablenotetext{f}{Radius of the \cb\ of an \ire.}
	\tablenotetext{g}{The emissivity at the radius {\bff larger than this radius} 
					from the center of gravity is assumed to be zero in {\bff the models.} 
					} 
	\end{center}
\end{table}
\end{landscape}

\clearpage

\begin{figure}
	\begin{center}
	\iffigure
	\includegraphics[bb = 0 0 1000 1000, scale = 0.42]{\dirfig 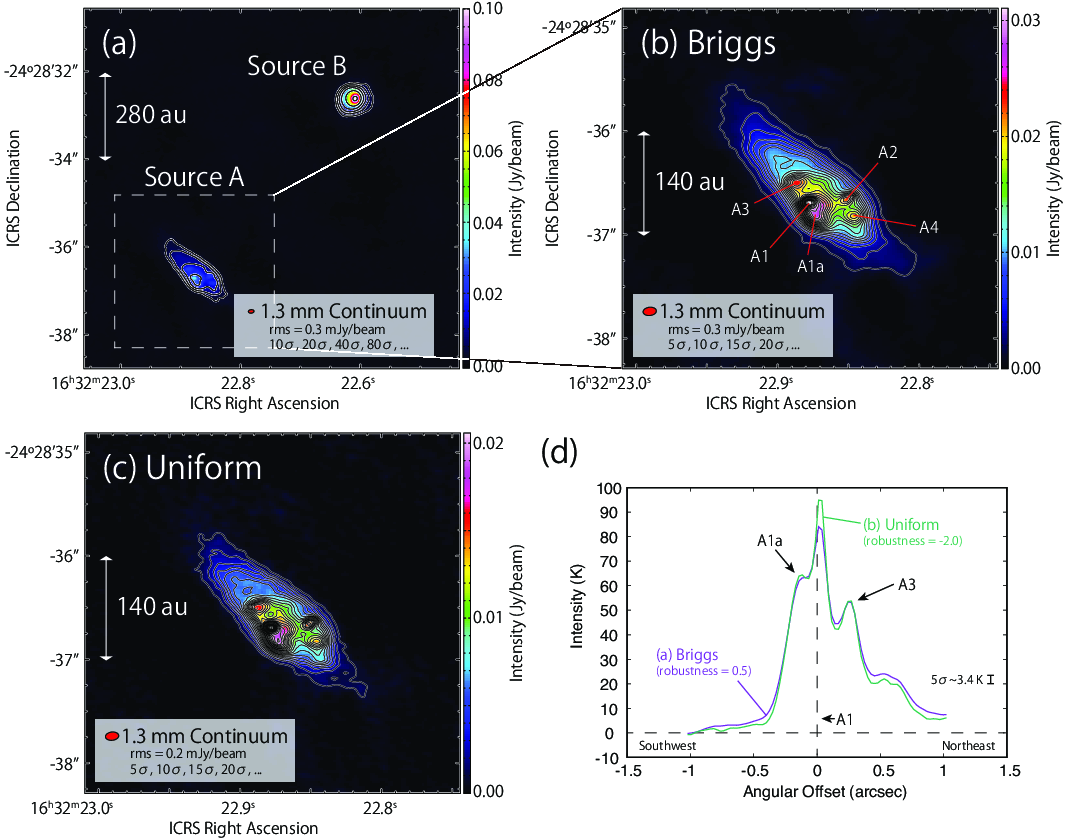}
	\fi
%
	\caption{\bff 1.3 mm continuum image. 
			Panels (a) and (b) are obtained with the Briggs weighting with a robustness parameter of 0.5, 
			while panel (c) is with the uniform weighting (i.e. robustness parameter of $-2.0$). 
			Contour levels in panel (a) are 10, 20, 40, 80, 160, and 320$\sigma$, 
			where the rms noise level is 0.3 \mJypb. 
			Contour levels in panels (b) and (c) are at intervals of 5$\sigma$ starting from 5$\sigma$, 
			where the rms noise level is 0.3 and 0.2 \mJypb, respectively. 
			Both the rms noise levels correspond to $\sim$0.7 K. 
			Color scales in panels (b) and (c) are set to be from 0 K to $\sim$70 K. 
			(d) Spatial profile of the continuum emission in panels (b) and (c). 
			The position axis is prepared along the line passing through 
			the continuum peak A1 and A1a. 
			\label{fig:cont}} 
	\end{center}
\end{figure}

\clearpage
\begin{figure}
	\begin{center}
	\iffigure
	\includegraphics[bb = 0 0 1000 1000, scale = 0.42]{\dirfig 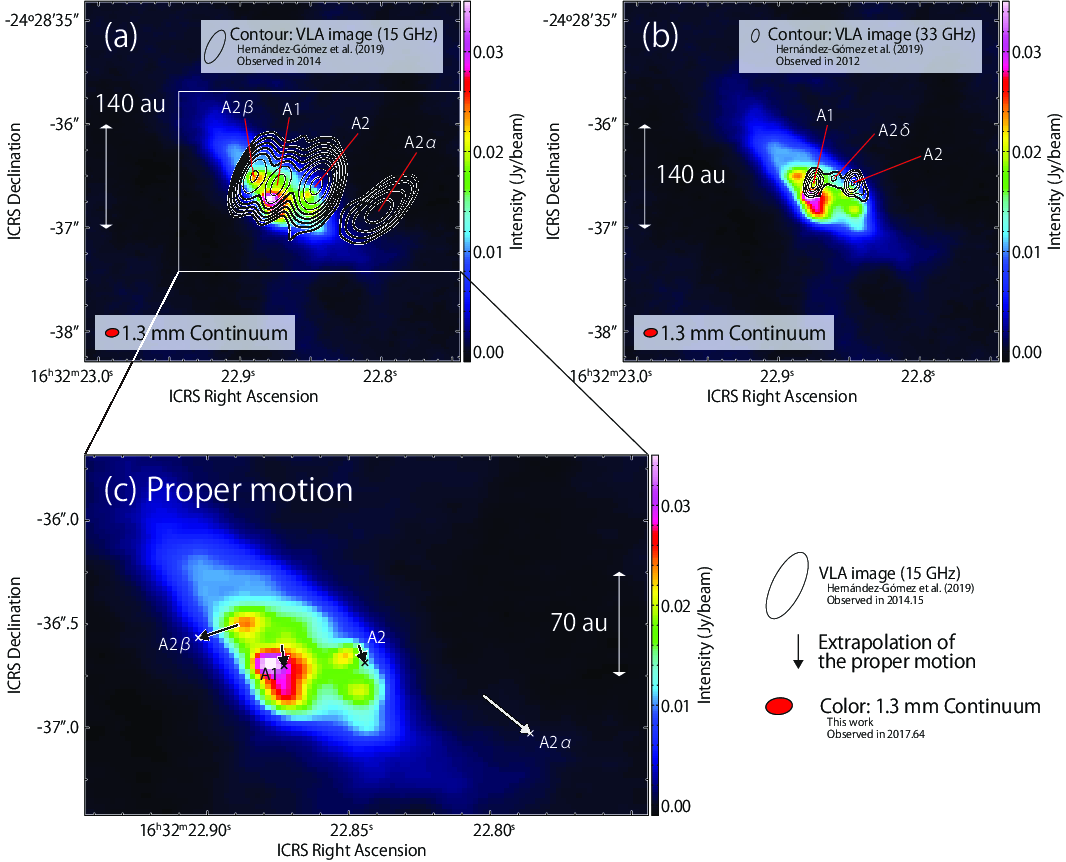} 
	\fi
	\caption{\bff (a, b) The continuum images at 15 GHz (a; contour) and 33 GHz (b; contour) 
			reported by \citet{HernandezGomez2019_contVLA} 
			overlaid on the 1.3 mm continuum image (color). 
			A beam size for each observation is depicted in the boxes in the corners. 
			(c) Extrapolation of the proper motion of the continuum peaks 
			(A1, A2, A2$\alpha$, and A2$\beta$) reported by \citet{HernandezGomez2019_contVLA}. 
			\label{fig:cont_cm}} 
	\end{center}
\end{figure}

\begin{landscape}
\begin{figure}
	\begin{center}
	\iffigure
	\includegraphics[bb = 0 0 1500 800, scale = 0.36]{\dirfig 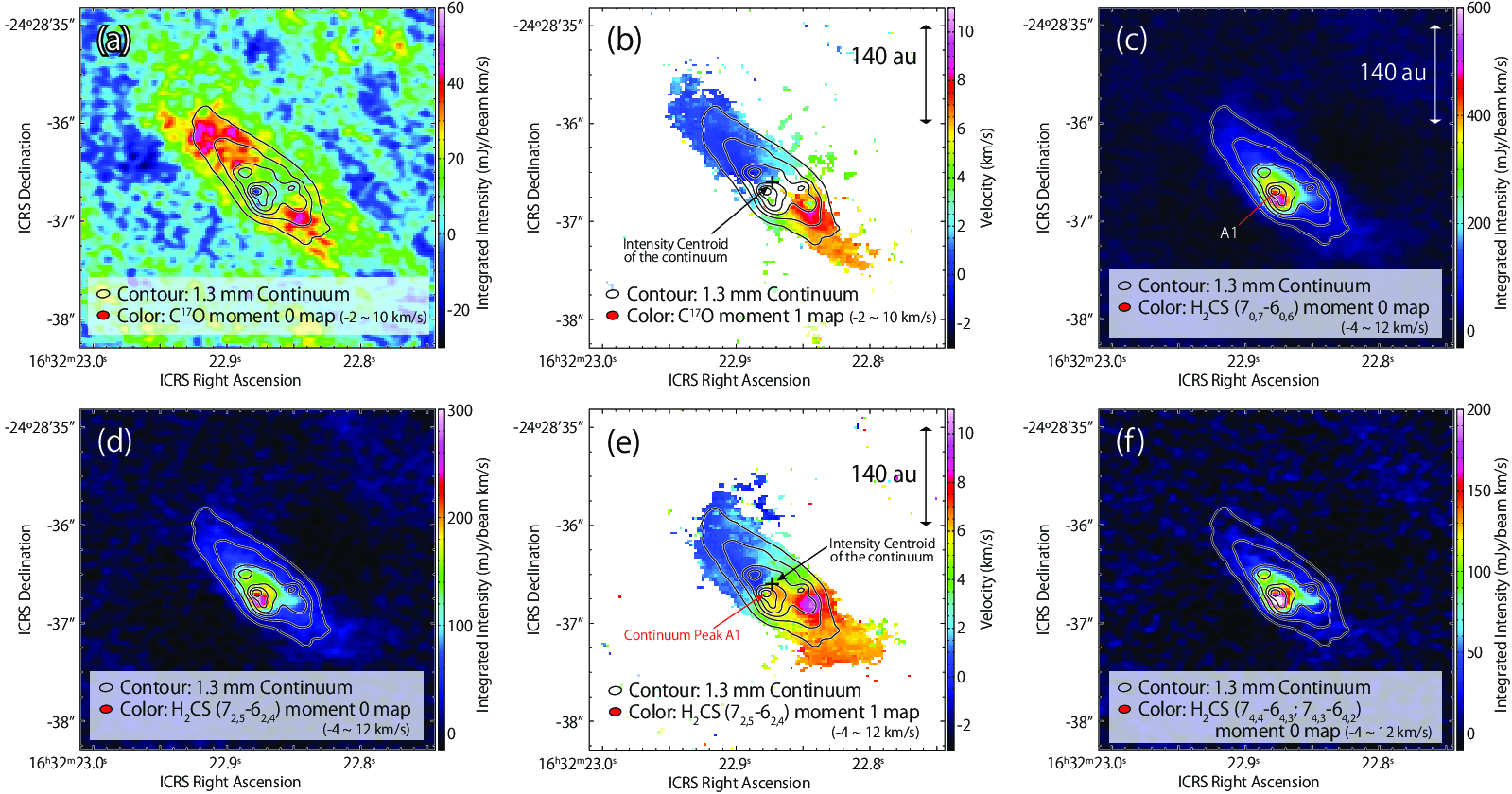}
	\fi
	\caption{Integrated intensity maps of the \CO\ (\co; a) and \TFA\ (c, d, f) lines, 
			and velocity maps (moment 1 maps) of the \CO\ (\co; b) and \TFA\ (\tfab; e) lines. 
			The velocity range for the integration is from $-2$ to $+10$ \kmps\ for the \CO\ line (a, b) 
			and from $-4$ to $+12$\kmps\ for the \TFA\ lines (c--f). 
			These velocity ranges correspond to the velocity-shift ranges 
			from $-6$ to $+6$ \kmps\ and from $-8$ to $+8$ \kmps\ with respect to the \sysV\ of Source A (\Vsys). 
			Contours in each panel represent the 1.3 mm continuum. 
			The contour levels are at intervals of 20$\sigma$ starting from 10$\sigma$, 
			where the rms noise level is 0.3 \mJypb. 
			In panels (b) and (e), 
			the {\bff moment 1} maps {\bff (color)} are 
			prepared for positions where the integrated intensity exceeds 3$\sigma$. 
			\label{fig:mom0}} 
	\end{center}
\end{figure}
\end{landscape}

\begin{landscape}
\begin{figure}
	\begin{center}
	\iffigure
	\includegraphics[bb = 0 0 1500 300, scale = 1.1]{\dirfigchan 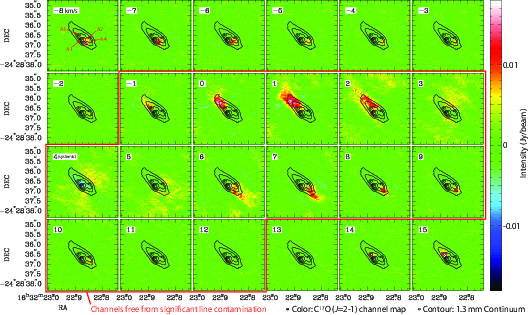}
	\fi
	\caption{Velocity channel maps of the \CO\ (\co) line. 
			{\bff Each panel is 
			prepared by averaging 
			the velocity range of 1 \kmps, 
			while the line cube has a channel width of 0.2 \kmps.}  
			\remContinuum \ 
			\remChan\ 
			The panels for the velocity range from $-8$ \kmps\ to $-2$ \kmps\ and from 13 to 15 \kmps\ are 
			affected by the contamination lines (see Section \ref{sec:results_co}). 
			\label{fig:chan_co}} 
	\end{center}
\end{figure}
\end{landscape}

\begin{landscape}
\begin{figure}
	\begin{center}
	\iffigure
	\includegraphics[bb = 0 0 1500 300, scale = 1.1]{\dirfigchan 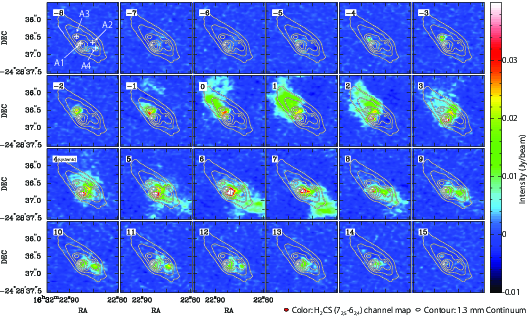}
	\fi
	\caption{Velocity channel maps of the \TFA\ (\tfab) line. 
			{\bff Each panel is 
			prepared by averaging 
			the velocity range of 1 \kmps, 
			while the line cube has a channel width of 0.2 \kmps.}  
			\remContinuum \ 
			\remChan
			\label{fig:chan_h2cs}} 
	\end{center}
\end{figure}
\end{landscape}

\begin{landscape}
\begin{figure}
	\begin{center}
	\iffigure
	\vspace*{-20pt}
	\includegraphics[bb = 0 0 1200 430, scale = 0.55]{\dirfig 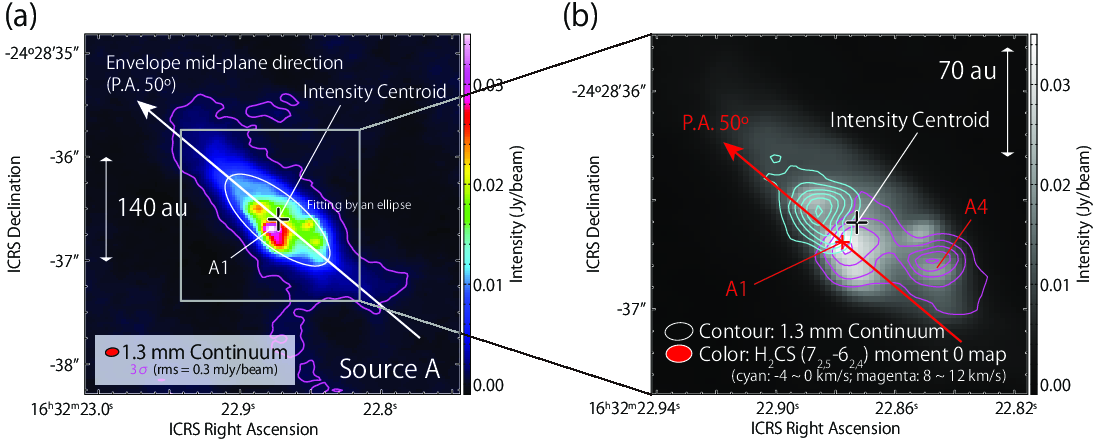}
	\fi
	\caption{(a) 1.3 mm continuum image. 
			The color map and black contours are the same as those in Figure \ref{fig:cont}(b). 
			The level for the magenta contour is 3$\sigma$. 
			{\bff 
			The white ellipse shows {\bff the result of} the fit of the continuum {\bff distribution} by an ellipse 
			by using the least-squares method weighted by the intensity at each position (see Section \ref{sec:disc_kin}). 
			The center of the ellipse shown by a  black cross is located at 
			$(\alpha_{\rm ICRS}, \delta_{\rm ICRS}) = (16^{\rm h}32^{\rm m}22\fs874, -24\degr28\arcmin36\farcs614)$, 
			which is obtained as the mean position of all the data points with $3\sigma$ ($0.9$ \mJypb; magenta contours) or higher of detection significance 
			weighted by the local intensity. 
			The sizes of the major and minor axes of the ellipse are 1\farcs25 ($\sim$170 au) and 0\farcs50 ($\sim$70 au), respectively. 
			}
			The white arrow represents the direction where the continuum emission extends 
			(`the envelope mid-plane direction'; see Section \ref{sec:disc_kin}). 
			The position angle of the {\bff arrow} is derived 
			to be $50\fdegr19\pm0\fdegr03$ 
			{\bff by the least-squares fit,} 
			where the error represents {\bff one} standard deviation. 
		(b) Integrated intensity maps of the \TFA\ (\tfab; contour) line 
			overlaid on the 1.3 mm continuum map (gray scale). 
			The velocity ranges for integration are from $-4$ to 0 \kmps\ and from $8$ to $12$ \kmps\ for 
			the blue-shifted component (cyan contours) and the red-shifted one (magenta contours), respectively. 
			These velocity ranges correspond to the velocity shift from $\pm4$ to $\pm8$ \kmps. 
			The contour levels are at intervals of 5$\sigma$ starting from 5$\sigma$, 
			where the rms noise level is 2.4 \mJypb\ \kmps. 
			The red cross shows the continuum peak position (A1 in Figure \ref{fig:cont}). 
			The red arrow 
			represents the direction of the velocity gradient of the \TFA\ line. 
			{\bff It} 
			is centered at the continuum peak A1 
			and {\bff its} 
			position angle is \PAire. 
			\label{fig:centroid}} 
	\end{center}
\end{figure}
\end{landscape}

\begin{figure}
	\begin{center}
	\iffigure
	\includegraphics[bb = 0 0 1000 1000, scale = 0.42]{\dirfig 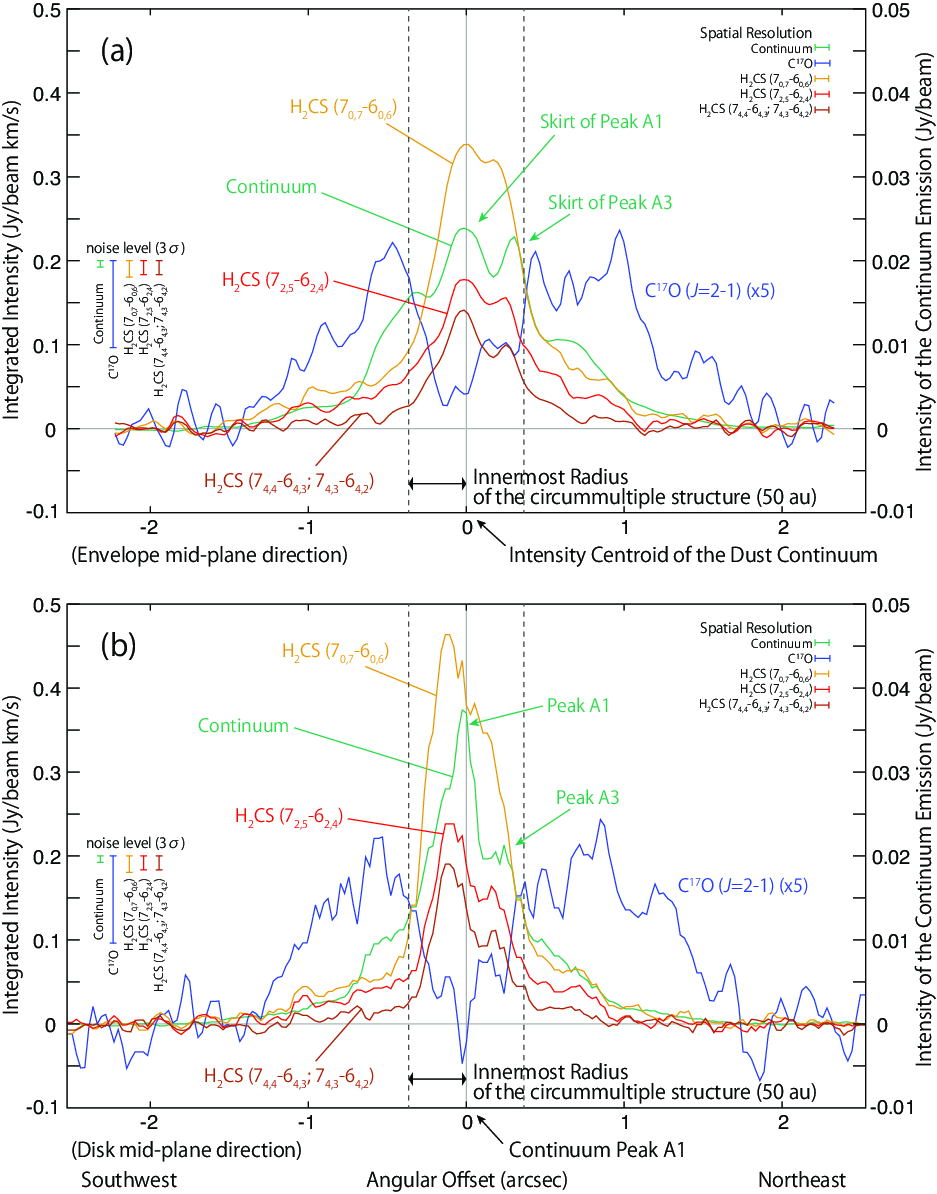}
	\fi
	\caption{The spatial profiles of the continuum emission 
			and the integrated intensities of the \CO\ and \TFA\ lines. 
			The position axis in panel (a) is along the envelope mid-plane direction, 
			which is shown by {\bff the} white arrow 
			{\bff passing through the intensity centroid of the continuum emission} 
			in Figure \ref{fig:centroid}(a) (Section \ref{sec:disc_kin}). 
			The position axis in panel (b) is along the disk mid-plane direction, 
			which is shown by {\bff the} red arrow 
			{\bff passing through the continuum peak A1} 
			in Figure \ref{fig:centroid}(b) (Section \ref{sec:disc_kin_tfa}). 
			\label{fig:spatialProfile}} 
	\end{center}
\end{figure}

\begin{figure}
	\begin{center}
	\iffigurePV
	\includegraphics[bb = 0 0 1700 1200, scale = 0.28]{\dirfig 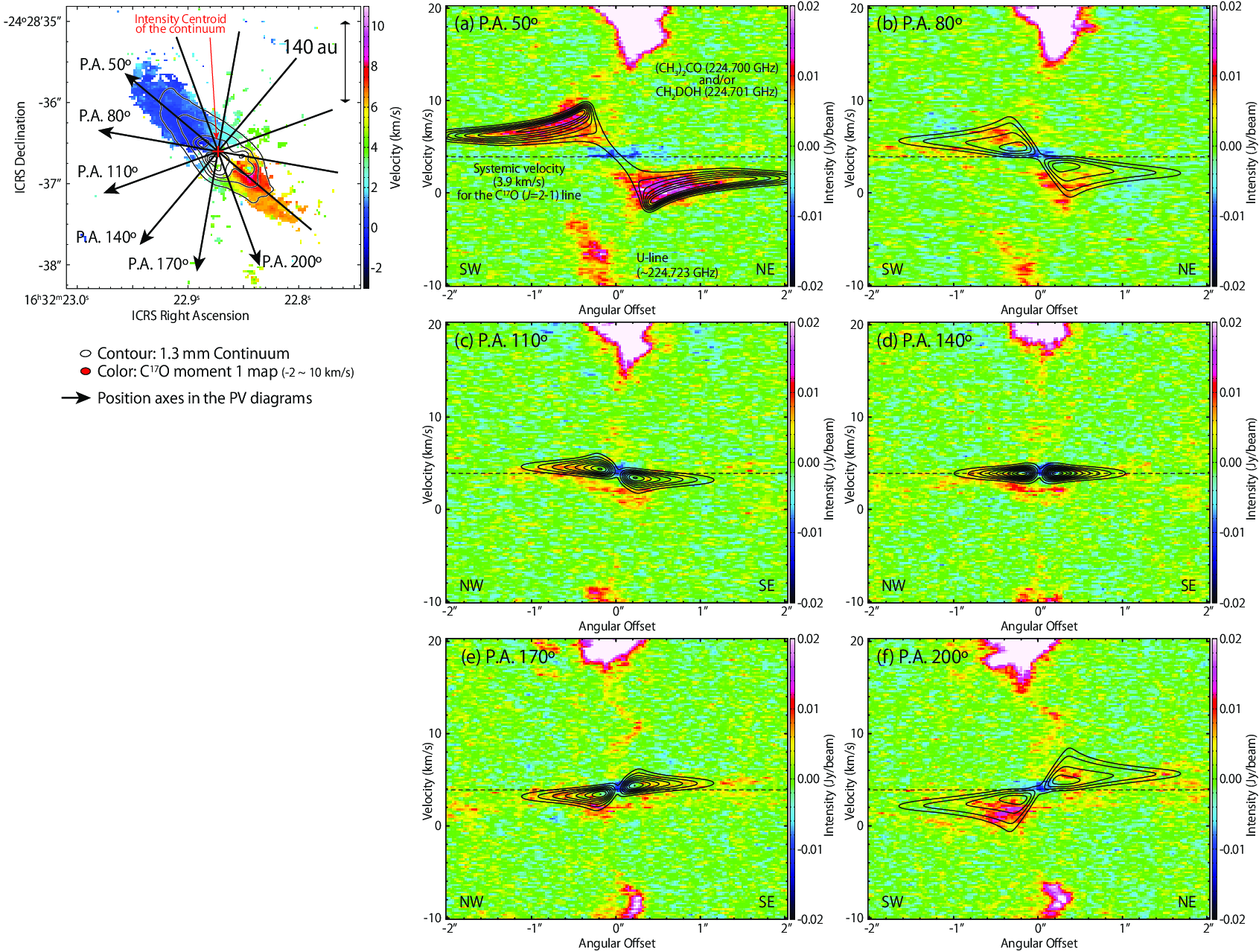}
	\fi
	\caption{{\bff Velocity map (moment 1 map; left panel) and position-velocity diagrams (a--f) of the \CO\ (\co) line. 
			The velocity map is the same as Figure \ref{fig:mom0}(b).
			Black arrows in the velocity map represent the position axes
			along which the PV diagrams are prepared. 
			They are centered at the intensity centroid of the continuum emission. 
			The contours in the PV diagrams represent 
			the results of the Keplerian disk model (contour) (see Section \ref{sec:disc_kin_co_Kep}).} 
			The physical parameters employed for the model are: 
			the central mass is \MdiskCO\ 
			and the \ia\ is \IdiskCO. 
			{\bff The emissivity in the model is assumed to be proportional to $r^{-1.5}$, 
			where $r$ denotes the distance from the center of gravity, 
			and is to be zero for $r <$ \RindiskCO\ and $r >$ \RoutdiskCO. 
			The scale height of the disk is assumed to increase as increasing the distance from the center of gravity 
			with the flared angle of 30\degr.} 
			Contour levels for the model result are at intervals of 10  \%\ of the peak intensity in the model cube. 
			\label{fig:PV_co-Kepler}} 
	\end{center}
\end{figure}

\begin{figure}
	\begin{center}
	\iffigurePV
	\includegraphics[bb = 0 0 1200 1250, scale = 0.4]{\dirfig 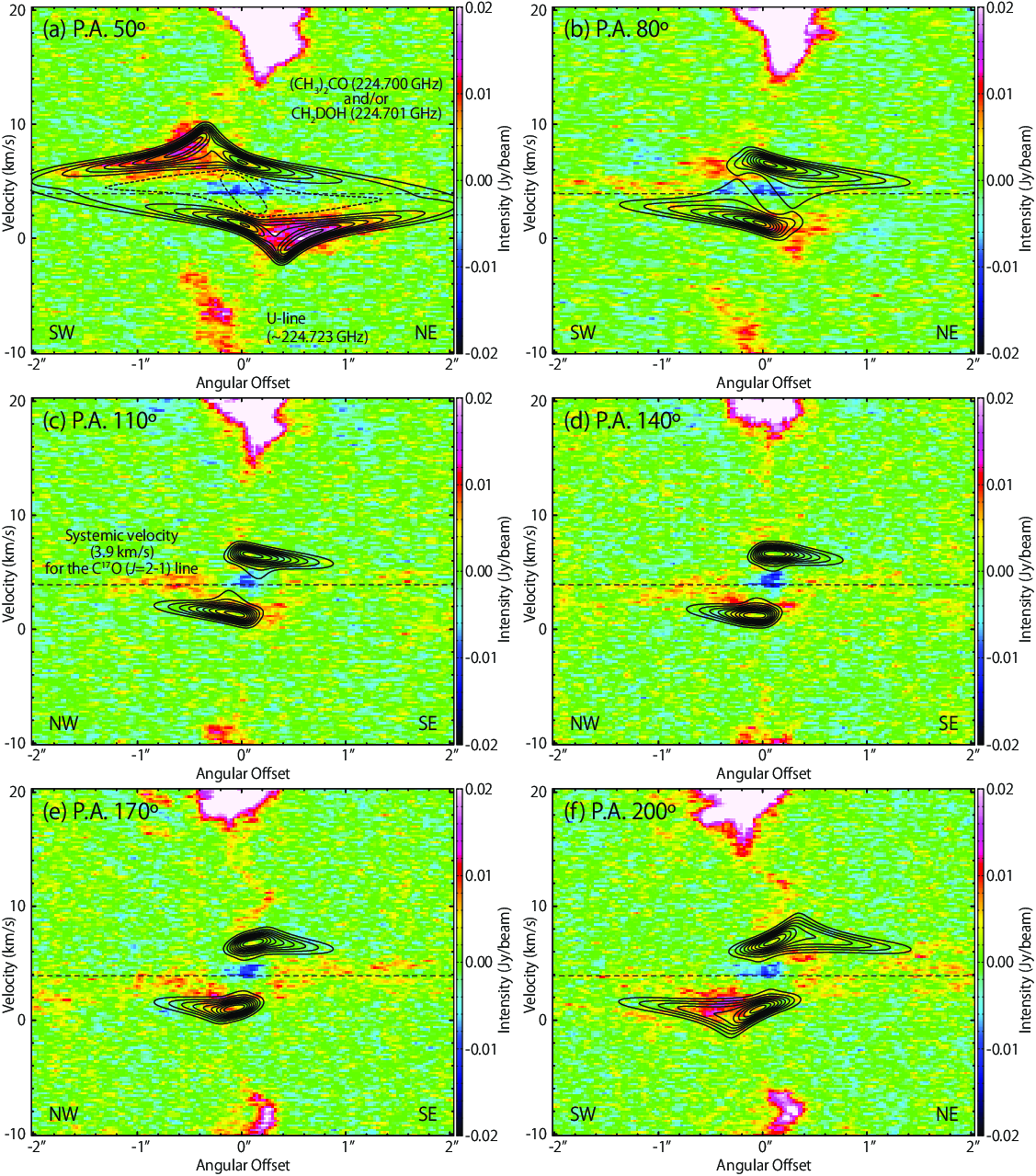}
	\fi
	\caption{Position-velocity diagrams of the \CO\ (\co; color) line and the results of the \ire\ model (contour) 
			{\bff for various position angles} 
			shown by arrows in {\bff Figure \ref{fig:PV_co-Kepler}(left panel),} 
			which are centered at the intensity centroid of the continuum emission. 
			The color maps are as the same as those in {\bff Figure \ref{fig:PV_co-Kepler}.} 
			The physical parameters employed for the model are: 
			the central mass is \Mire, 
			the \ia\ is \Iire, 
			and the radius of the \cb\ is \CBire. 
			The emissivity in the model is assumed to be proportional to $r^{-1.5}$, 
			where $r$ denotes the distance from the center of gravity, 
			and is to be zero for $r <$ \CBire\ and $r >$ \RoutireCO. 
			The scale height of the envelope is assumed to increase as increasing the distance from the center of gravity 
			with the flared angle of 30\degr. 
			Contour levels for the model result are at intervals of 10  \%\ of the peak intensity in the model cube. 
			The dashed contours in {\bff panel (a)} represent the dip toward the central position. 
			\label{fig:PV_co-IRE}} 
	\end{center}
\end{figure}

\begin{figure}
	\begin{center}
	\vspace*{-20pt}
	\iffigurePV
	\includegraphics[bb = 0 0 1700 1200, scale = 0.28]{\dirfig 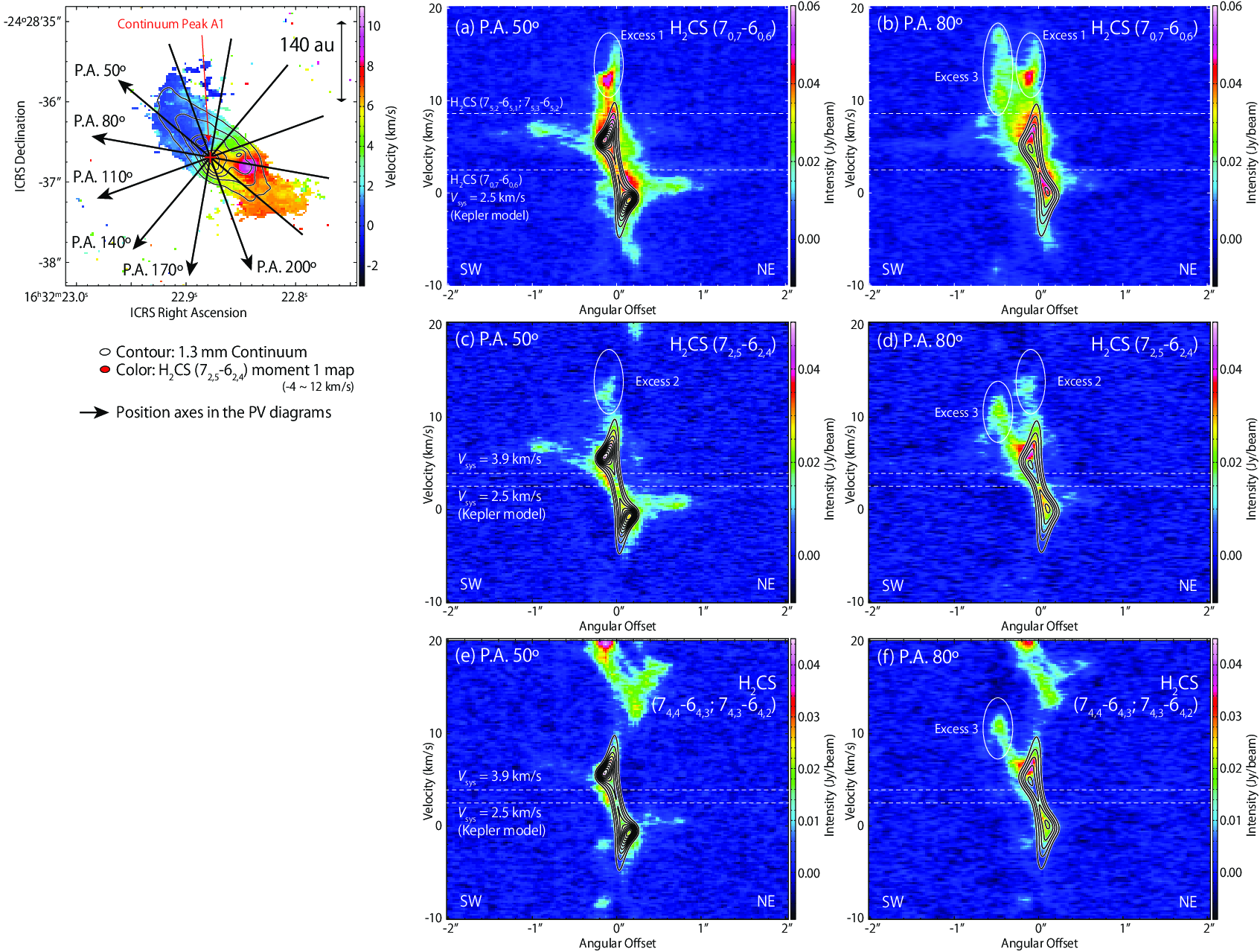}
	\fi
	\caption{Position-velocity diagrams of 
			{\bff the \tfaa\ (a, b), \tfab\ (c, d), and \tfad\ (e, f) lines of \TFA\ along 
			the directions with P.A. of 50\degr\ and 80\degr\ and} 
			the results of the Keplerian disk model (contour). 
			{\bff The left panel shows the velocity map of the \TFA\ (\tfab) line, 
			which is the same in Figure \ref{fig:mom0}(e). 
			Black arrows in the velocity map represent the position axes 
			along which the PV diagrams 
			{\bff of this figure and Figure \ref{fig:PV_h2cs-744-IRE}} 
			are prepared.} 
			{\bff They} are centered at the continuum peak A1. 
			{\bff The physical parameters employed for the model are as follows:
			the central mass is \Mdisk, 
			and the \ia\ is \Idisk\ {\bff (Table \ref{tb:modelparams}).} 
			The center of gravity for the model is assumed to be at the continuum peak A1, 
			and the systemic velocity is assumed to be \Vsysdisk, 
			instead of \Vsys\ employed for the \ire\ model 
			for the circummultiple \desys. 
			The emissivity in the model is assumed to be proportional to $r^{-1.5}$, 
			where $r$ denotes the distance from the protostar, 
			and is to be zero for $r >$ \Routdisk. 
			The scale height of the emitting region is assumed to increase as increasing the distance from the protostar 
			with the flared angle of 30\degr.} 
			The intensity in the models is convolved with the synthesized beam size for 
			{\bff each molecular line} (Table \ref{tb:lines}) 
			and the Gaussian line profile with the intrinsic line width (FWHM) of 1 \kmps. 
			Contour levels for the model result are at intervals of 10  \%\ of the peak intensity in the model cube. 
			\label{fig:PV_h2cs-707-725-744-Kep}} 
	\end{center}
\end{figure}

\begin{figure}
	\begin{center}
	\iffigurePV
	\includegraphics[bb = 0 0 1700 1200, scale = 0.28]{\dirfig 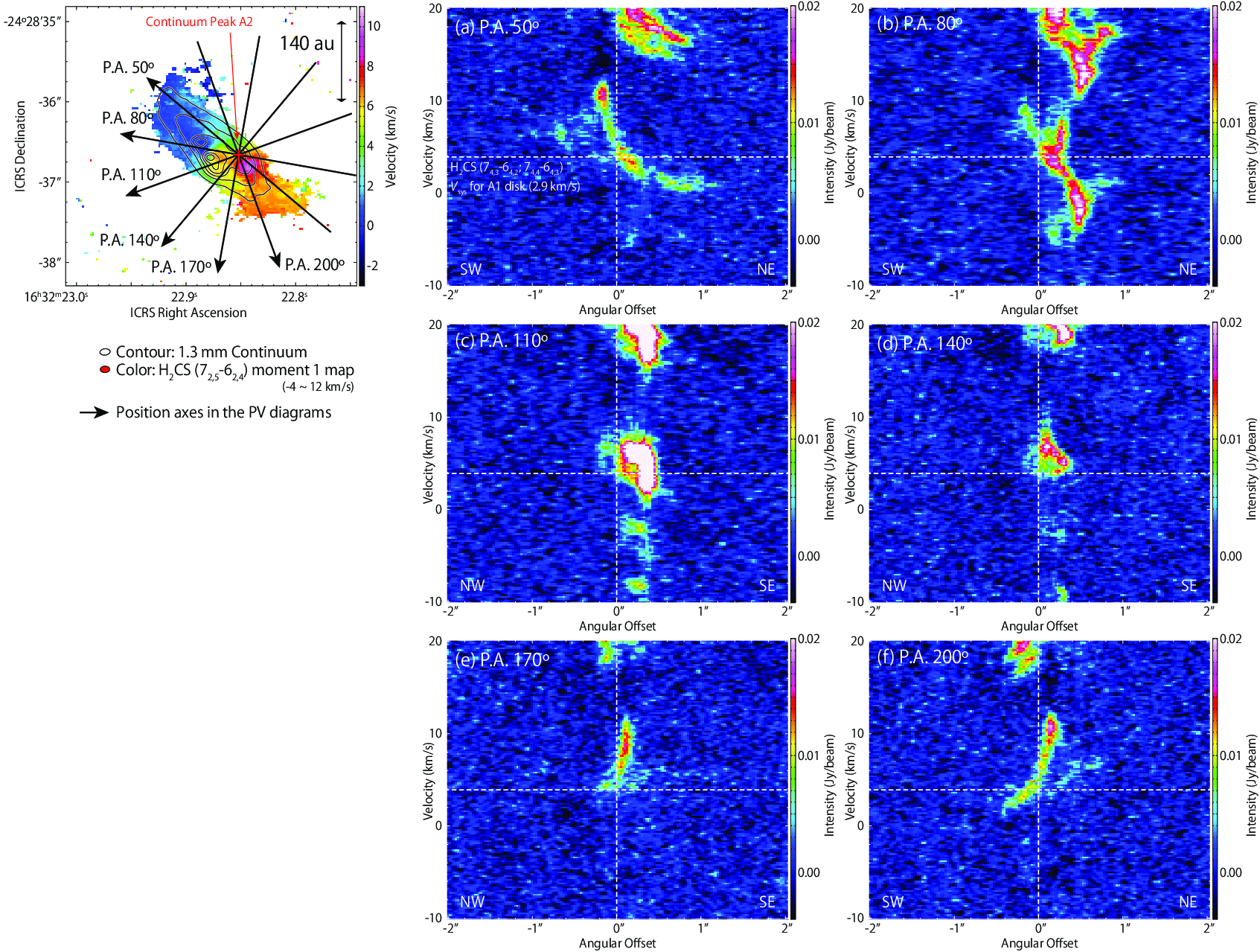}
	\fi
	\caption{\bff Position-velocity diagrams of the \TFA\ (\tfad; color) line. 
			The left panel shows the velocity maps of the \TFA\ (\tfab) line, 
			which is the same in Figure \ref{fig:mom0}(e). 
			Black arrows in the velocity map represent the position axes 
			along which the PV diagrams are prepared, 
			which are centered at the continuum peak A2. 
			Vertical white dashed lines represent the position of A2. 
			Horizontal white dashed lines represent 
			the \sysV\ assumed for the disk associated to A1 (\Vsysdisk), 
			for reference. 
			\label{fig:PV_h2cs-744_A2}} 
	\end{center}
\end{figure}

\begin{landscape}
\begin{figure}
	\begin{center}
	\iffigure
	\includegraphics[bb = 0 0 1200 430, scale = 0.55]{\dirfig 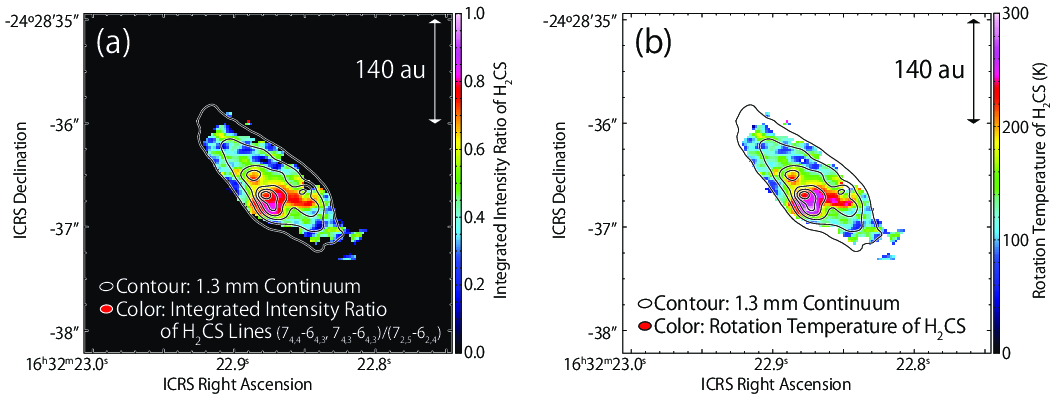}
	\fi
	\caption{(a) Map of the integrated intensity ratio of the \TFA\ (\tfad) line relative to the \TFA\ (\tfab) line. 
			The velocity range for the integrations is from $-4$ to $+12$ \kmps, 
			where the \sysV\ of Source A is \Vsys. 
			(b) Map of the rotation temperature derived from the integrated intensity ratio 
			between the two \TFA\ lines shown in panel (a). 
			The optically thin conditions for the two \TFA\ lines and 
			the LTE condition are assumed (see Section \ref{sec:disc_temp}). 
			\remContinuum 
			\label{fig:mom0_h2cs-Trot}} 
	\end{center}
\end{figure}
\end{landscape}

\begin{figure}
	\begin{center}
	\iffigure
	\includegraphics[bb = 0 0 900 1200, scale = 0.42]{\dirfig 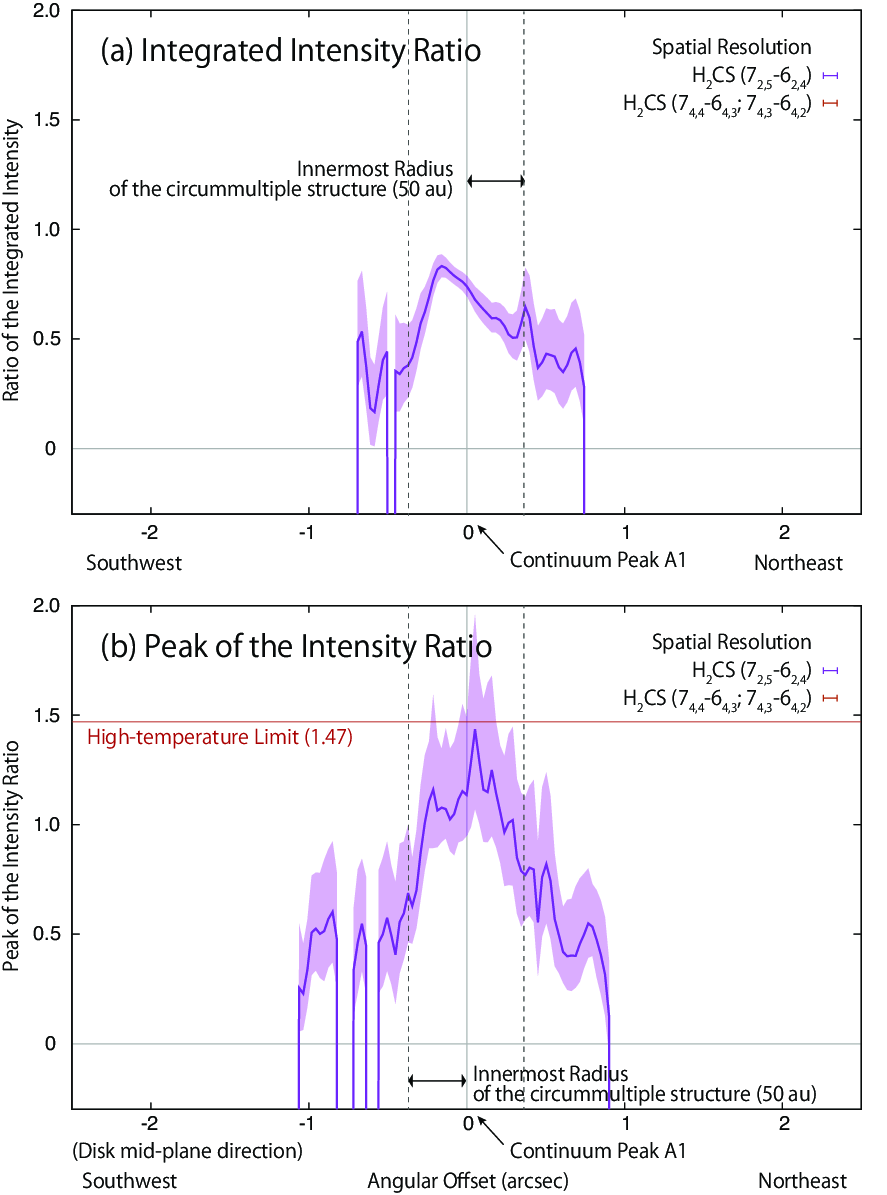}
	\fi
	\caption{(a) Spatial profile of the ratio of the integrated intensity 
			of the \TFA (\tfad) line relative to the \TFA\ (\tfab) line. 
			(b) Spatial profile of the peak value of the intensity ratio 
			of the \TFA (\tfad) line relative to the \TFA\ (\tfab) line. 
			The position axes are along the disk mid-plane direction, 
			which is shown by a red arrow in Figure \ref{fig:centroid}(b) (Section \ref{sec:disc_kin_tfa_A1}). 
			The colored {\bff areas} represent the error ranges; 
			for instance, 
			the lower limit for the integrated intensity ratio in panel a 
			is taken as ${\displaystyle \left(I_4 - \sigma \right) / \left(I_2 + \sigma \right)}$, 
			where $I_4$ and $I_2$ are the integrated intensities of the \TFA\ (\tfad) and (\tfab) lines 
			and $\sigma$ stands for their rms noise level (6 \mJypb\ \kmps). 
			The rms noise level for the {\bff integrated intensity used in panel a is 6 \mJypb\ \kmps, 
			while that for the} line intensities used in panel b is 2 \mJypb. 
			Red horizontal line in panel b represents the high-temperature limit (1.47 for $T=\infty$). 
			\label{fig:spatialProfile_intensity}} 
	\end{center}
\end{figure}

\begin{figure}
	\begin{center}
	\iffigure
	\includegraphics[bb = 0 0 900 1200, scale = 0.42]{\dirfig 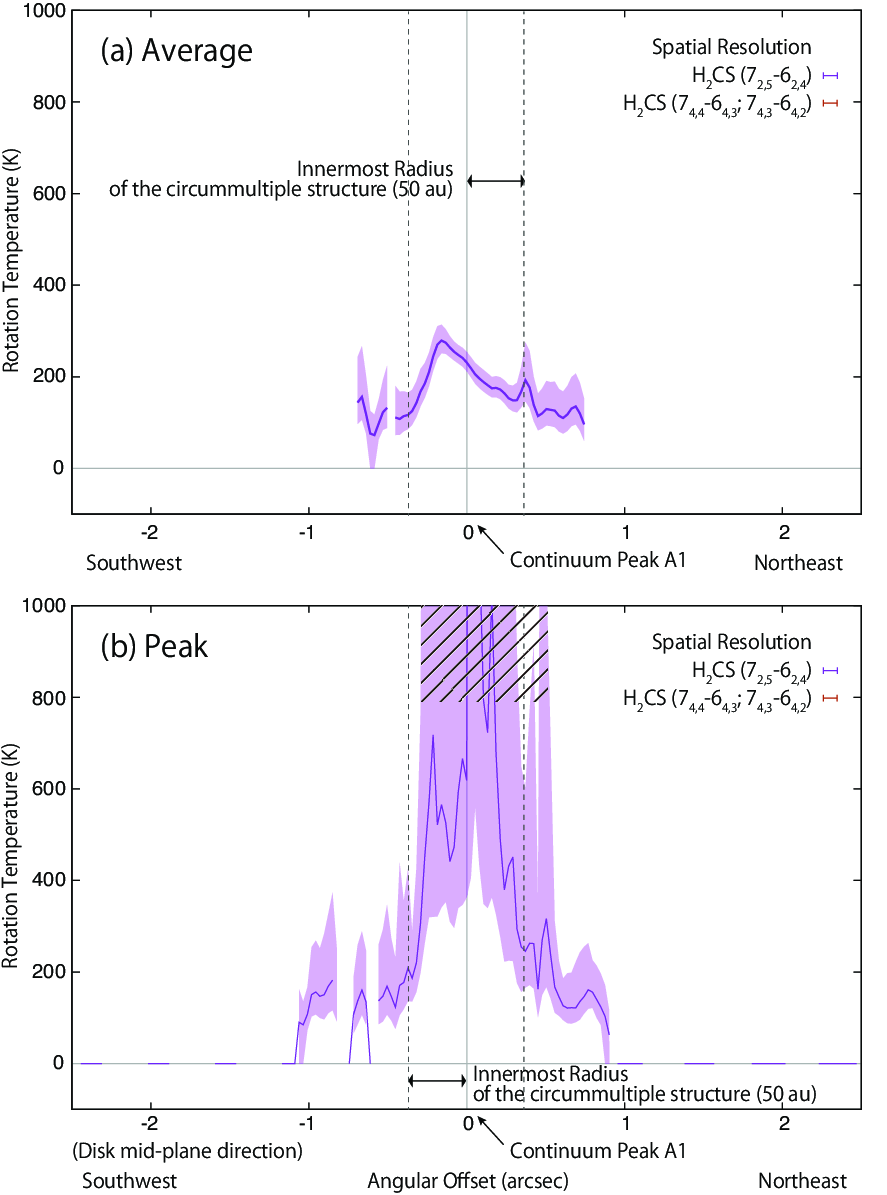}
	\vspace*{-15pt}
	\fi
	\caption{Spatial profiles of the rotation temperature of \TFA\ obtained from the \TFA\ (\tfab) and (\tfad) lines. 
			(a) The rotation temperature is calculated 
			{\bff from} the ratio of the integrated intensities between the two \TFA\ lines (Figure \ref{fig:spatialProfile_intensity}a). 
			This can be regarded as the averaged value along the line of sight. 
			(b) The rotation temperature is calculated 
			{\bff from the peak intensity ratio over the velocity range for each position} (Figure \ref{fig:spatialProfile_intensity}b). 
			This can be regarded as the peak rotation temperature along the line of sight. 
			The position axes are along the disk mid-plane direction, 
			which is shown by a red arrow in Figure \ref{fig:centroid}(b) (Section \ref{sec:disc_kin_tfa_A1}). 
			The colored {\bff areas} represent the error ranges, 
			where the lower and upper limit are calculated from those in Figure \ref{fig:spatialProfile_intensity}. 
			The region with the halftone dot meshing represents 
			where the intensity ratio is higher than 1.2. 
			The very high temperatures ($> 790$ K) in this region is artificial 
			since the intensity ratio is {\bff close to} the high-temperature limit (1.47 for $T=\infty$) (Figure \ref{fig:spatialProfile_intensity}). 			
			\label{fig:spatialProfile_Trot}} 
	\end{center}
\end{figure}

\begin{landscape}
\begin{figure}
	\begin{center}
	\iffigure
	\includegraphics[bb = 0 0 1500 300, scale = 1.1]{\dirfigchan 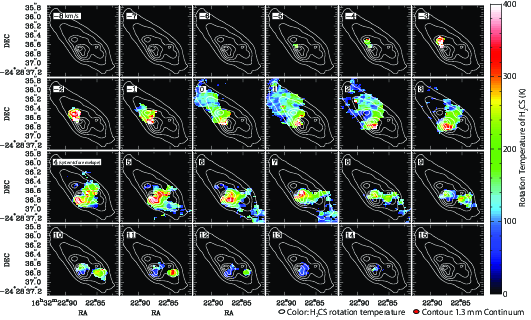}
	\fi
	\caption{Velocity channel maps of the rotation temperature of \TFA. 
			The \TFA\ (\tfab) and (\tfad) lines are employed for the calculation. 
			{\bff Each panel is 
			prepared by averaging 
			the velocity range of 1 \kmps, 
			while the line cube has a channel width of 0.2 \kmps.}  
			The optically thin conditions and the LTE condition are assumed. 
			\remContinuum \ 
			The center velocity for each panel is shown in the upper-left corner, 
			where the \sysV\ of the envelope is \Vsys. 
			\label{fig:chan_h2cs-Trot}} 
	\end{center}
\end{figure}
\end{landscape}

\begin{landscape}
\begin{figure}
	\begin{center}
	\iffigure
	\includegraphics[bb = 0 0 1200 430, scale = 0.55]{\dirfig 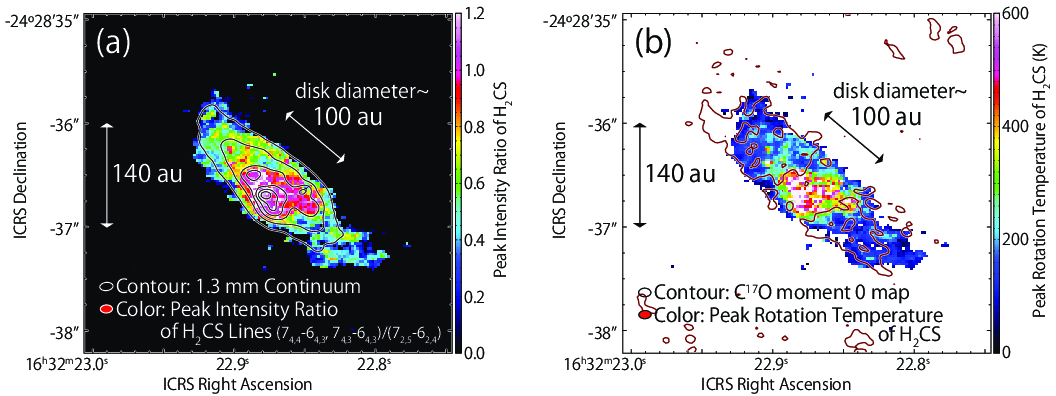}
	\fi
	\caption{(a) Map of the maximum value of the intensity ratio of the \TFA\ (\tfad) line relative to the \TFA\ (\tfab) line 
			along the velocity channels in the cube data (see Figure \ref{fig:chan_h2cs-Trot}). 
			\remContinuum \ 
			(b) Map of the maximum value of the rotation temperature of \TFA\ 
			along the velocity channels in the cube data. 
			Contours represent the integrated intensity of the \CO\ line shown in Figure \ref{fig:mom0}(a). 
			\remContour{3}{3}{7} 
			The \CO\ emission is weaker for the higher temperature region. 
			\label{fig:mom8_h2cs-Trot}} 
	\end{center}
\end{figure}
\end{landscape}

\begin{figure}
	\begin{center}
	\iffigure
	\includegraphics[bb = 0 0 1000 200, scale = 0.8]{\dirfig 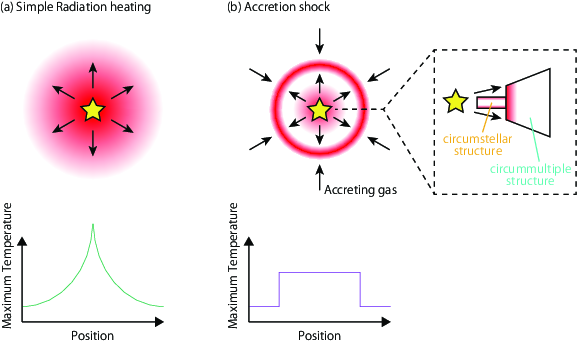}
	\fi
	\caption{Schematic illustrations for the gas temperature distribution in \iras\ Source A. 
			(a) The gas temperature distribution is expected to be a power-law of the distance from the protostar, 
			if a simple radiation heating by the protostar is considered. 
			(b) If the infalling gas causes an accretion shock, 
			the gas temperature can locally rise there. 
			Moreover, the physical structure of the \desys\ may cause a local rise of the gas temperature; 
			if the infalling envelope gas is stagnated and piled up vertically 
			near the transition zone, 
			it can be efficiently heated by the radiation from the protostar. 
			The expected profile of the maximum gas temperature along the \los\ is 
			schematically shown for each situation, 
			assuming that the system has an almost edge-on configuration. 
			This profile 
			{\bff can be qualitatively compared to} 
			{\bff Figures \ref{fig:spatialProfile_Trot}(b)} 
			and \ref{fig:mom8_h2cs-Trot}(b) in the observation. 
			\label{fig:scheme_temp}} 
	\end{center}
\end{figure}

\appendix
\setcounter{table}{0}
\setcounter{figure}{0}
\makeatletter
	\def\thetable{A\the\c@table}
	\def\thefigure{A\the\c@figure}
\makeatother

\begin{landscape}
\begin{table}
	\begin{center}
	\caption{
			Reduced $\chi^2$ Test on the Keplerian Model for the PV diagram of \CO\ \tablenotemark{a} 
			\label{tb:chi2_C17O-Kep}}
	\begin{tabular}{cccccccccc}
	\hline \hline 
	& \multicolumn{8}{c}{Central Mass (\Msun)} \\ 
	Inclination Angle\tablenotemark{b} (\degr) & 0.5 & 1.0 & 1.5 & 2.0 & 2.5 & 3.0 & 3.5 & 4.0 \\ \hline
        90 & 5.84 & 5.14 & \fbox{4.32} & \fbox{4.32} & 5.78 & 7.80 & 9.42 & 10.64 \\
        80 & 5.82 & 5.16 & \fbox{4.33} & \fbox{4.17} & 5.47 & 7.45 & 9.25 & 10.53 \\
        70 & 5.61 & 4.65 & \fbox{3.80} & \doublebox{3.38} & \fbox{4.27} & 6.69 & 10.92 & 16.58 \\
        60 & 5.65 & 4.85 & \fbox{4.33} & \fbox{3.96} & 4.54 & 5.52 & 6.28 & 6.41 \\
        50 & 5.84 & 5.25 & 4.71 & \fbox{4.30} & \fbox{4.22} & 4.69 & 5.44 & 6.03 \\
        40 & 6.10 & 5.56 & 5.11 & 4.70 & 4.54 & \fbox{4.26} & \fbox{4.30} & 4.60 \\
        30 & 6.42 & 5.94 & 5.62 & 5.49 & 5.07 & 4.83 & 4.73 & 4.56 \\
        \hline 
	\end{tabular}
	\tablenotetext{a}{The \ire\ models are compared with the PV diagram of {\bff the} \CO\ (\co) line. 
					The PV diagram is prepared 
					along the envelope mid-plane direction (P.A. \PAdiskCO), 
					which is centered at the intensity centroid of the continuum emission. 
					The pixels with the angular offset from $0\arcsec$ to $+2\arcsec$ and the velocity from \VCOblueInf\ to \VCOblueSup  
					are taken into account. 
					2480 pixels are used for the calculation. 
					The minimum value is highlighted by a double box. 
					Single boxes highlight values whose difference from the minimum value 
					is lower than 1.} 
	\tablenotetext{b}{\incRem.}
	\end{center}
\end{table}
\end{landscape}

\begin{landscape}
\begin{table}
	\begin{center}
	\caption{
			Reduced $\chi^2$ Test on the Infalling-Rotating Envelope Model for the PV diagram of \CO\ \tablenotemark{a} 
			\label{tb:chi2_C17O-IRE}}
	\begin{tabular}{cccccccccccc}
	\hline \hline 
	& \multicolumn{10}{c}{Central Mass (\Msun)} \\ 
	Inclination Angle\tablenotemark{b} (\degr) & 0.2 & 0.4 & 0.6 & 0.8 & 1.0 & 1.2 & 1.4 & 1.6 & 1.8 & 2.0 \\ \hline
        90 & 5.77 & 4.45 & \fbox{3.15} & \fbox{2.37} & \fbox{2.24} & \fbox{2.67} & 3.51 & 4.90 & 6.72 & 8.78 \\
        80 & 5.79 & 4.49 & 3.19 & \fbox{2.36} & \doublebox{2.18} & \fbox{2.55} & 3.34 & 4.67 & 6.42 & 8.41 \\
        70 & 5.73 & 4.77 & 3.85 & 3.24 & \fbox{3.02} & 3.39 & 4.16 & 5.25 & 10.59 & 16.56 \\
        60 & 5.91 & 5.25 & 4.60 & 4.11 & 3.85 & 4.12 & 4.82 & 5.85 & 7.19 & 8.81 \\
        50 & 6.09 & 5.55 & 5.10 & 4.71 & 4.36 & 4.12 & 4.17 & 4.51 & 4.98 & 5.56 \\
        40 & 6.30 & 5.83 & 5.51 & 5.20 & 4.86 & 4.58 & 4.40 & 4.20 & 4.13 & 4.25 \\
        30 & 6.48 & 6.24 & 5.88 & 5.66 & 5.46 & 5.21 & 4.96 & 4.83 & 4.62 & 4.47 \\
	\hline 
	\end{tabular}
	\tablenotetext{a}{The \ire\ models are compared with the PV diagram of {\bff the} \CO\ (\co) line. 
					The PV diagram is prepared 
					along the envelope mid-plane direction (P.A. \PAire), 
					which is centered at the intensity centroid of the continuum emission. 
					The pixels with the angular offset from $0\arcsec$ to $+2\arcsec$ and the velocity from \VCOblueInf\ to \VCOblueSup  
					are taken into account. 
					2480 pixels are used for the calculation. 
					The minimum value is highlighted by a double box. 
					Single boxes highlight values whose difference from the minimum value 
					is lower than 1.} 
	\tablenotetext{b}{\incRem.}
	\end{center}
\end{table}
\end{landscape}

\begin{landscape}
\begin{table}
	\begin{center}
	\caption{
			Reduced $\chi^2$ Test on the Keplerian Model for the PV diagram of \TFA\ \tablenotemark{a} 
			\label{tb:chi2_h2cs744-Kep}}
	\begin{tabular}{ccccccccccc}
	\hline \hline 
	& \multicolumn{10}{c}{Central Mass (\Msun)} \\ 
	Inclination Angle\tablenotemark{b} (\degr) & 0.1 & 0.2 & 0.3 & 0.4 & 0.5 & 0.6 & 0.7 & 0.8 & 0.9 & 1.0 \\ \hline 
        90 & 12.75 & 11.21 & 9.03 & 8.51 & 9.12 & 10.34 & 11.52 & 12.67 & 13.72 & 14.50 \\ 
        80 & 12.62 & 10.92 & 8.73 & 8.14 & 8.77 & 10.02 & 11.43 & 12.68 & 13.88 & 14.78 \\ 
        70 & 12.41 & 10.34 & 8.08 & \fbox{7.34} & \fbox{7.91} & 9.25 & 10.92 & 12.70 & 14.29 & 15.74 \\ 
        60 & 12.17 & 10.41 & 8.17 & \doublebox{7.07} & \fbox{7.21} & 8.32 & 10.05 & 12.18 & 14.34 & 16.41 \\ 
        50 & 12.48 & 11.11 & 9.39 & \fbox{7.87} & \fbox{7.29} & \fbox{7.56} & 8.56 & 10.13 & 12.09 & 14.32 \\ 
        40 & 13.03 & 11.95 & 11.02 & 9.83 & 8.65 & \fbox{7.95} & \fbox{7.77} & \fbox{8.04} & 8.76 & 9.81 \\ 
        30 & 13.69 & 12.85 & 12.22 & 11.74 & 11.29 & 10.59 & 9.82 & 9.12 & 8.63 & 8.32 \\ 
        20 & 14.12 & 13.73 & 13.32 & 12.96 & 12.62 & 12.39 & 12.26 & 12.12 & 12.05 & 11.65 \\ 
        10 & 14.61 & 14.32 & 14.18 & 14.02 & 13.91 & 13.82 & 13.68 & 13.64 & 13.47 & 13.44 \\ 
        \hline 
	\end{tabular}
	\tablenotetext{a}{The Keplerian models are compared with the two PV diagrams of \TFA\ (\tfad) line. 
					The PV diagrams are prepared 
					along the disk mid-plane direction (P.A. \PAdisk) and the direction perpendicular to it (P.A. \PAdiskperp), 
					which are centered at the continuum peak A1. 
					The pixels with the angular offset from $-0\farcs4$ to $+0\farcs4$ and the velocity from \VTFAInf\ to \VTFASup 
					are taken into account. 
					5676 pixels are used for the calculation. 
					The minimum value is highlighted by a double box. 
					Single boxes highlight values whose difference from the minimum value 
					is lower than 1.} 
	\tablenotetext{b}{\incRem.}
	\end{center}
\end{table}
\end{landscape}

\begin{table}
	\begin{center}
	\caption{Reduced $\chi^2$ Test on the Infalling-Rotating Envelope Model for the PV diagram of \TFA\ \tablenotemark{a} 
			\label{tb:chi2_h2cs744-IRE}}
	\begin{tabular}{ccccccc}
	\hline \hline 
	& & \multicolumn{5}{c}{Central Mass (\Msun)} \\ 
	Radius of the CB\tablenotemark{b} (au) & Inclination Angle\tablenotemark{c} (\degr) & 0.10 & 0.15 & 0.20 & 0.25 & 0.30 \\ \hline
	5 & 90 & 14.49 & 12.92 & 12.46 & 13.59 & 15.94 \\
	& 80 & 13.50 & 11.85 & 11.56 & 12.64 & 14.90 \\
	& 70 & 12.73 & 11.07 & 10.58 & 11.29 & 12.96 \\
	& 60 & 12.95 & 11.30 & 10.51 & 10.76 & 11.73 \\
	& 50 & 13.86 & 12.82 & 11.82 & 11.51 & 11.80 \\ \hline 
	10 & 90 & 12.86 & 10.95 & 10.40 & 11.53 & 13.77 \\
	& 80 & 12.31 & 10.22 & \fbox{9.62} & 10.55 & 12.53 \\
	& 70 & 11.94 & \fbox{9.84} & \doublebox{8.98} & \fbox{9.48} & 10.96 \\
	& 60 & 12.50 & 10.75 & \fbox{9.44} & \fbox{9.37} & 10.19 \\
	& 50 & 13.81 & 12.79 & 11.29 & 10.51 & 10.58 \\ \hline 
	20 & 90 & 13.15 & 11.22 & \fbox{9.80} & 10.04 & 11.76 \\
	& 80 & 13.06 & 11.35 & \fbox{9.84} & 9.99 & 11.54 \\
	& 70 & 13.63 & 12.14 & 10.58 & 10.31 & 11.52 \\
	& 60 & 14.63 & 13.85 & 12.33 & 11.55 & 11.91 \\
	& 50 & 14.76 & 14.36 & 13.77 & 12.89 & 12.36 \\ 
 	\hline 
	\end{tabular}
	\tablenotetext{a}{The \ire\ models are compared with the two PV diagrams of \TFA\ (\tfad) line. 
					The PV diagrams are prepared 
					along the disk mid-plane direction (P.A. \PAdisk) and the direction perpendicular to it (P.A. \PAdiskperp), 
					which are centered at the continuum peak A1. 
					The pixels with the angular offset from $-0\farcs4$ to $+0\farcs4$ and the velocity from \VTFAInf\ to \VTFASup 
					are taken into account. 
					5676 pixels are used for the calculation. 
					The minimum value is highlighted by a double box. 
					Single boxes highlight values whose difference from the minimum value 
					is lower than 1.} 
	\tablenotetext{b}{The radius of the \cb.} 
	\tablenotetext{c}{\incRem.}
	\end{center}
\end{table}

\begin{landscape}
\begin{figure}
	\begin{center}
	\iffigurePV
	\vspace*{-20pt}
	\includegraphics[bb = 0 0 1700 1300, scale = 0.3]{\dirfig 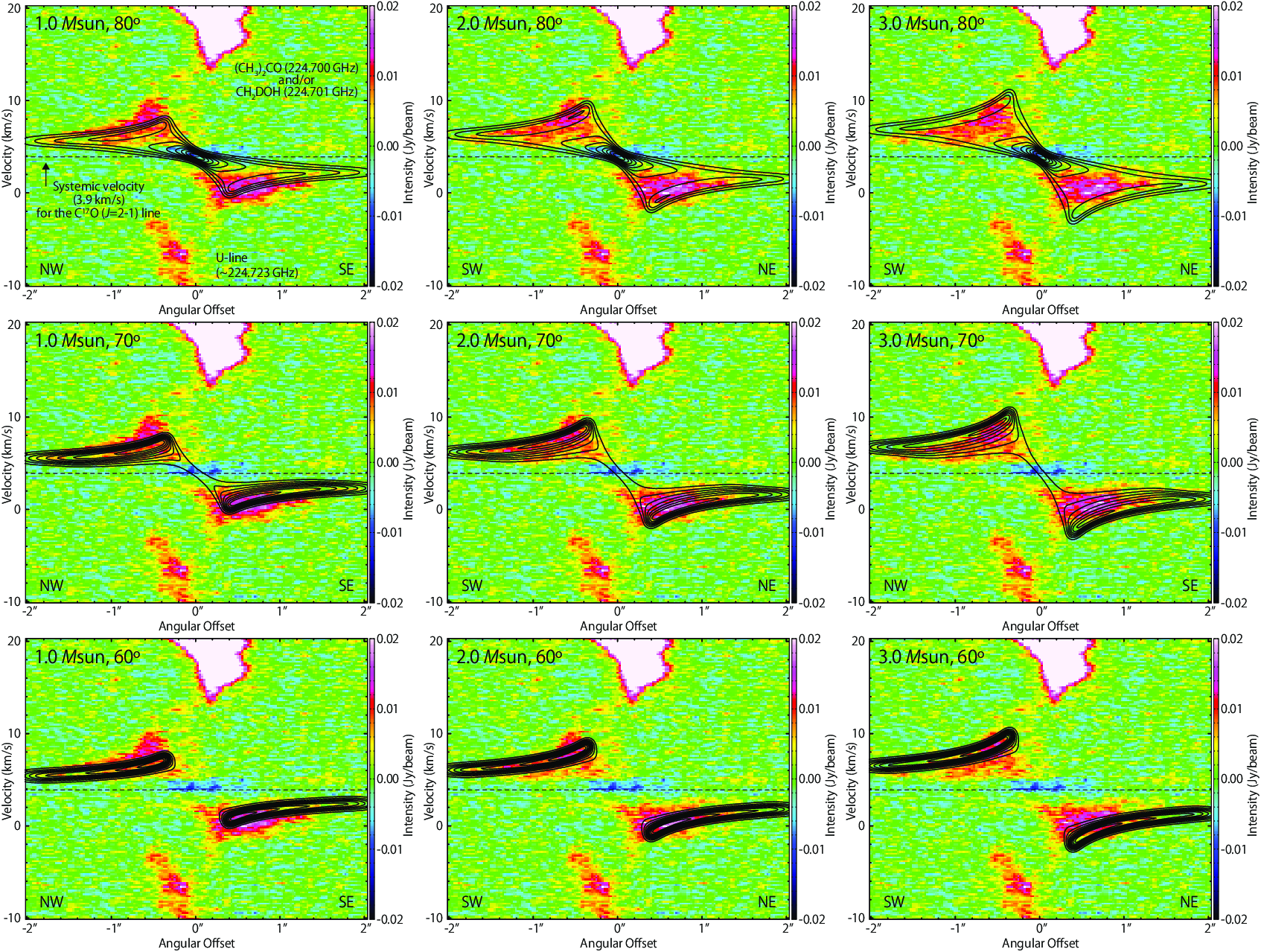}
	\fi
	\caption{Position-velocity diagrams of the \CO\ (\co; color) line and the results of the Keplerian disk model (contour). 
			Position axes in the panels are shown by arrows in {\bff Figure \ref{fig:PV_co-Kepler},} 
			which are centered at the intensity centroid of the continuum emission. 
			The color maps are as the same as those in {\bff Figure \ref{fig:PV_co-Kepler}.} 
			The physical parameters employed for the model are: 
			the central mass is 1.0, 2.0, and 3.0 \Msun, 
			and the \ia\ is 60\degr, 70\degr, and 80\degr. 
			{\bff Other details of the model are given in the caption of Figure \ref{fig:PV_co-Kepler}.} 
			Contour levels for the model result are at intervals of 10  \%\ of the peak intensity in the model cube. 
			\label{fig:PV_co-Kepler_MI}} 
	\end{center}
\end{figure}
\end{landscape}

\begin{landscape}
\begin{figure}
	\begin{center}
	\iffigurePV
	\vspace*{-20pt}
	\includegraphics[bb = 0 0 1700 1300, scale = 0.3]{\dirfig 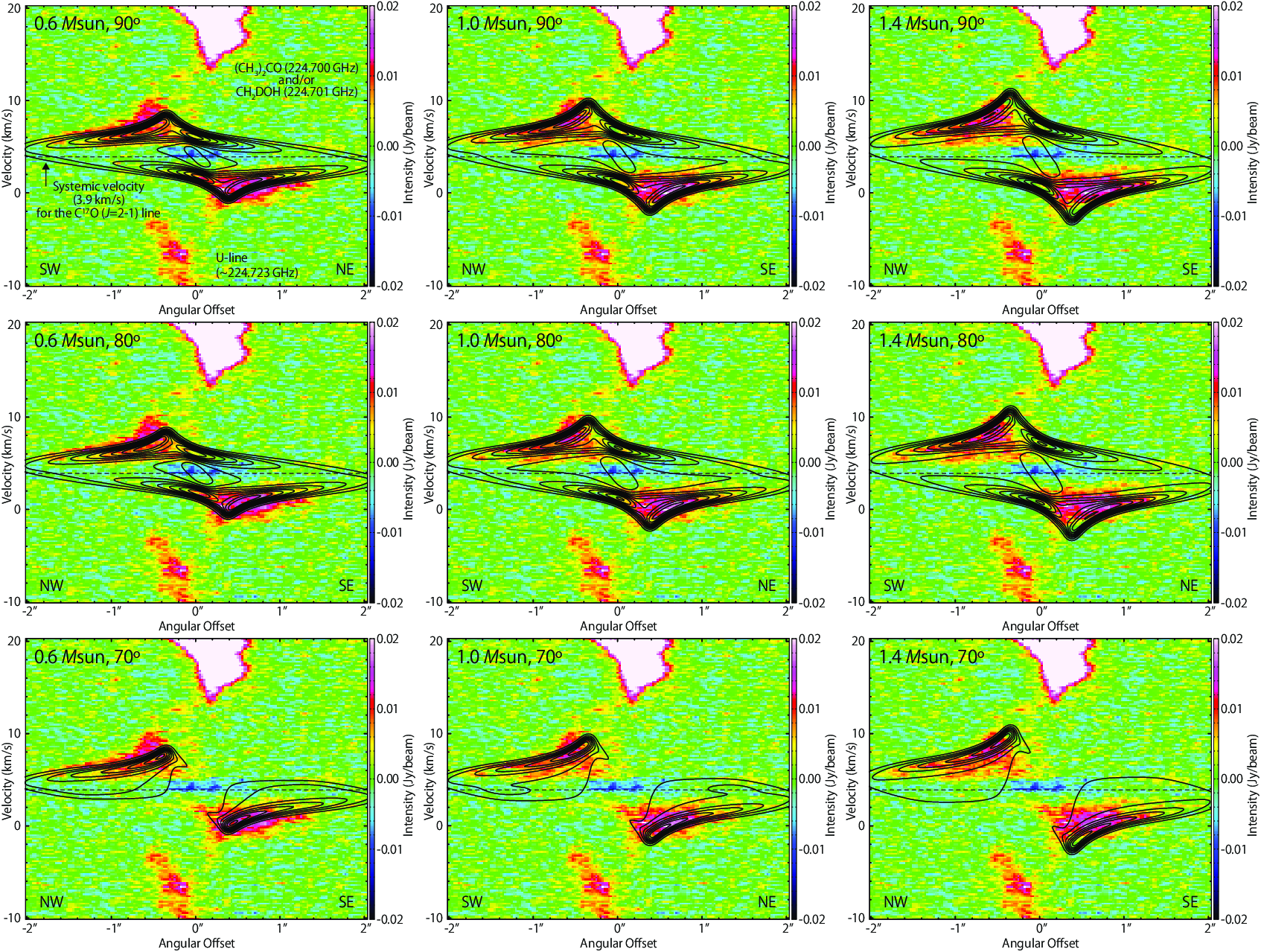}
	\fi
	\caption{Position-velocity diagrams of the \CO\ (\co; color) line and the results of the \ire\ model (contour). 
			Position axes in the panels are shown by arrows in {\bff Figure \ref{fig:PV_co-Kepler},} 
			which are centered at the intensity centroid of the continuum emission. 
			The color maps are as the same as those in {\bff Figure \ref{fig:PV_co-Kepler}.} 
			The physical parameters employed for the model are: 
			the central mass is 0.6, 1.0, and 1.4 \Msun, 
			the \ia\ is 70\degr, 80\degr, and 90\degr, 
			and the radius of the \cb\ is \CBire. 
			{\bff Other details of the model are given in the caption of Figure \ref{fig:PV_co-IRE}.} 
			Contour levels for the model result are at intervals of 10  \%\ of the peak intensity in the model cube. 
			The dashed contours in panels (a) and (b) represent the dip toward the central position. 
			\label{fig:PV_co-IRE_MI}} 
	\end{center}
\end{figure}
\end{landscape}

\begin{figure}
	\begin{center}
	\iffigurePV
	\vspace*{-20pt}
	\includegraphics[bb = 0 0 1000 1300, scale = 0.42]{\dirfig 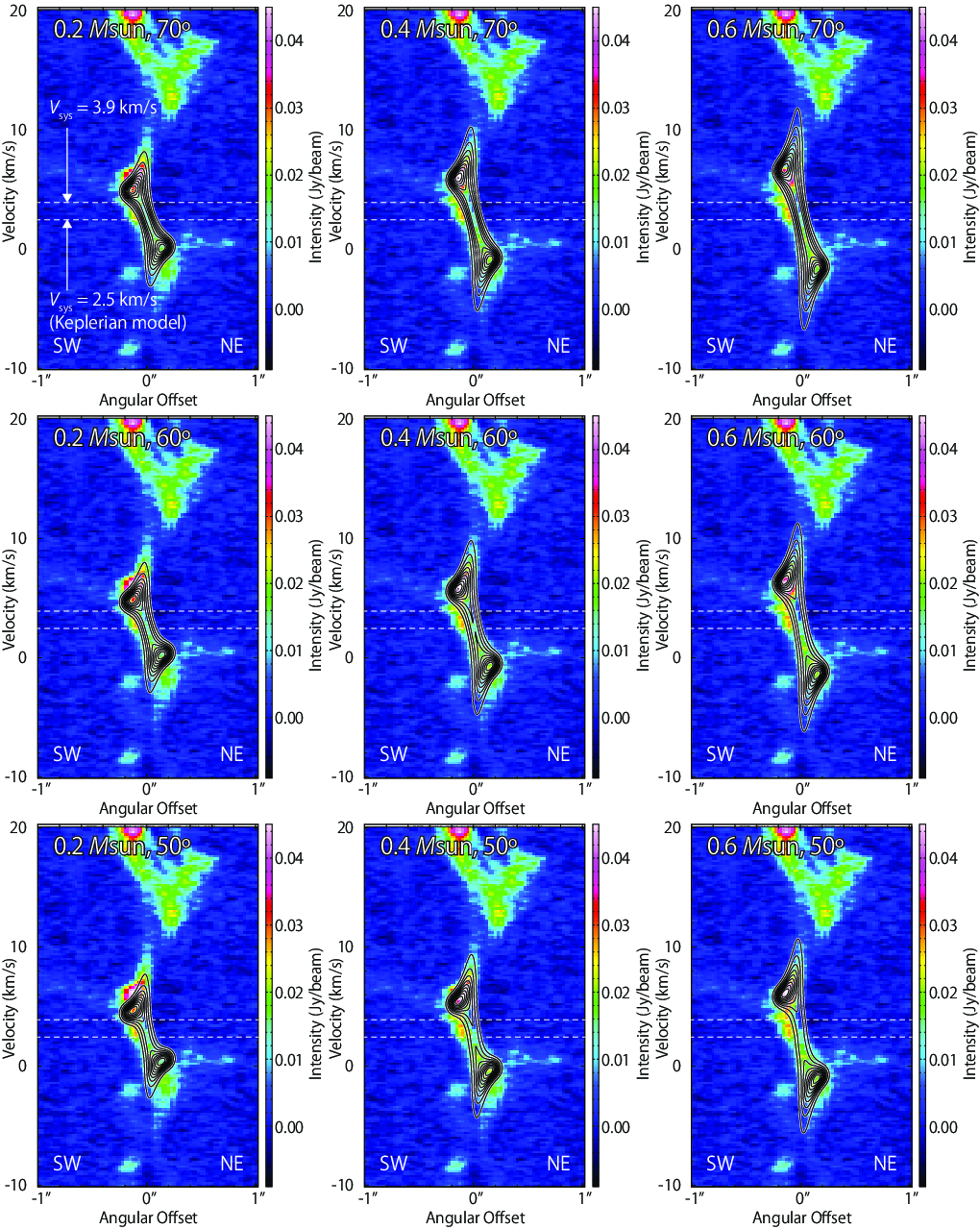}
	\vspace*{-10pt}
	\fi
	\caption{Position-velocity diagrams of the \TFA\ (\tfad; color) line and the Keplerian disk model (contour). 
			Position axis in the panels is prepared along the disk mid-plane direction (P.A. 50\degr) 
			shown by {\bff the red arrow in Figure \ref{fig:centroid}(b),} 
			which is centered at the continuum peak A1. 
			Black contours represent the Keplerian model results. 
			The physical parameters employed for the model are: 
			the central mass is 0.2, 0.4, and 0.6 \Msun, 
			and the \ia\ is 50\degr, 60\degr, and 70\degr. 
			{\bff Other details of the model are given in the caption of Figure \ref{fig:PV_h2cs-707-725-744-Kep}.} 
			Contour levels for the model result are at intervals of 10  \%\ of the peak intensity in the model cube. 
			\label{fig:PV_h2cs-744-Kep_MI-PA050}} 
	\end{center}
\end{figure}

\begin{figure}
	\begin{center}
	\iffigurePV
	\vspace*{-20pt}
	\includegraphics[bb = 0 0 1000 1300, scale = 0.42]{\dirfig 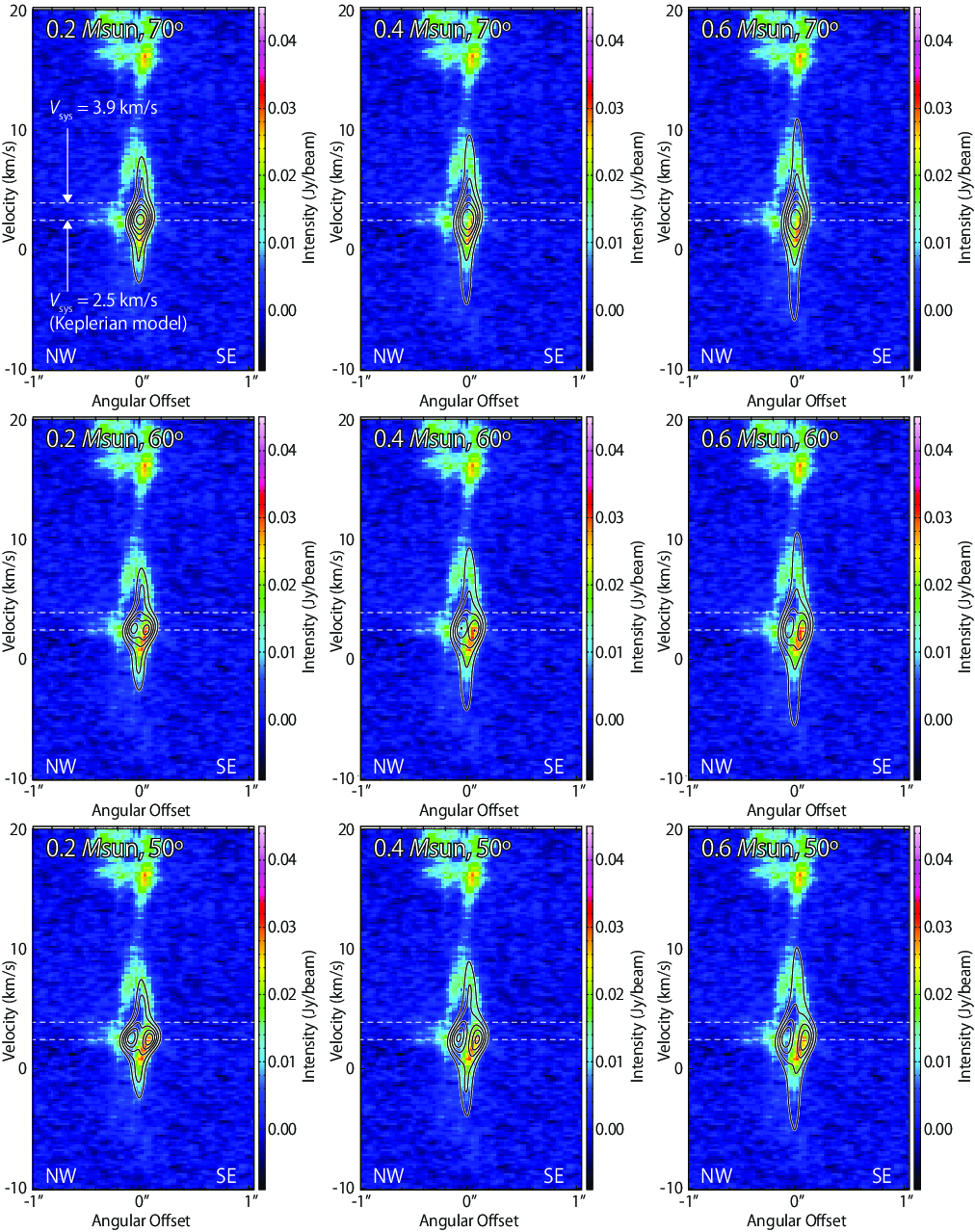}
	\vspace*{-10pt}
	\fi
	\caption{{\bff Same as Figure \ref{fig:PV_h2cs-744-Kep_MI-PA050}, 
			but along the direction perpendicular to the disk mid-plane direction (P.A. 140\degr).} 
			\label{fig:PV_h2cs-744-Kep_MI-PA140}} 
	\end{center}
\end{figure}

\begin{figure}
	\begin{center}
	\iffigurePV
	\includegraphics[bb = 0 0 1200 1250, scale = 0.4]{\dirfig 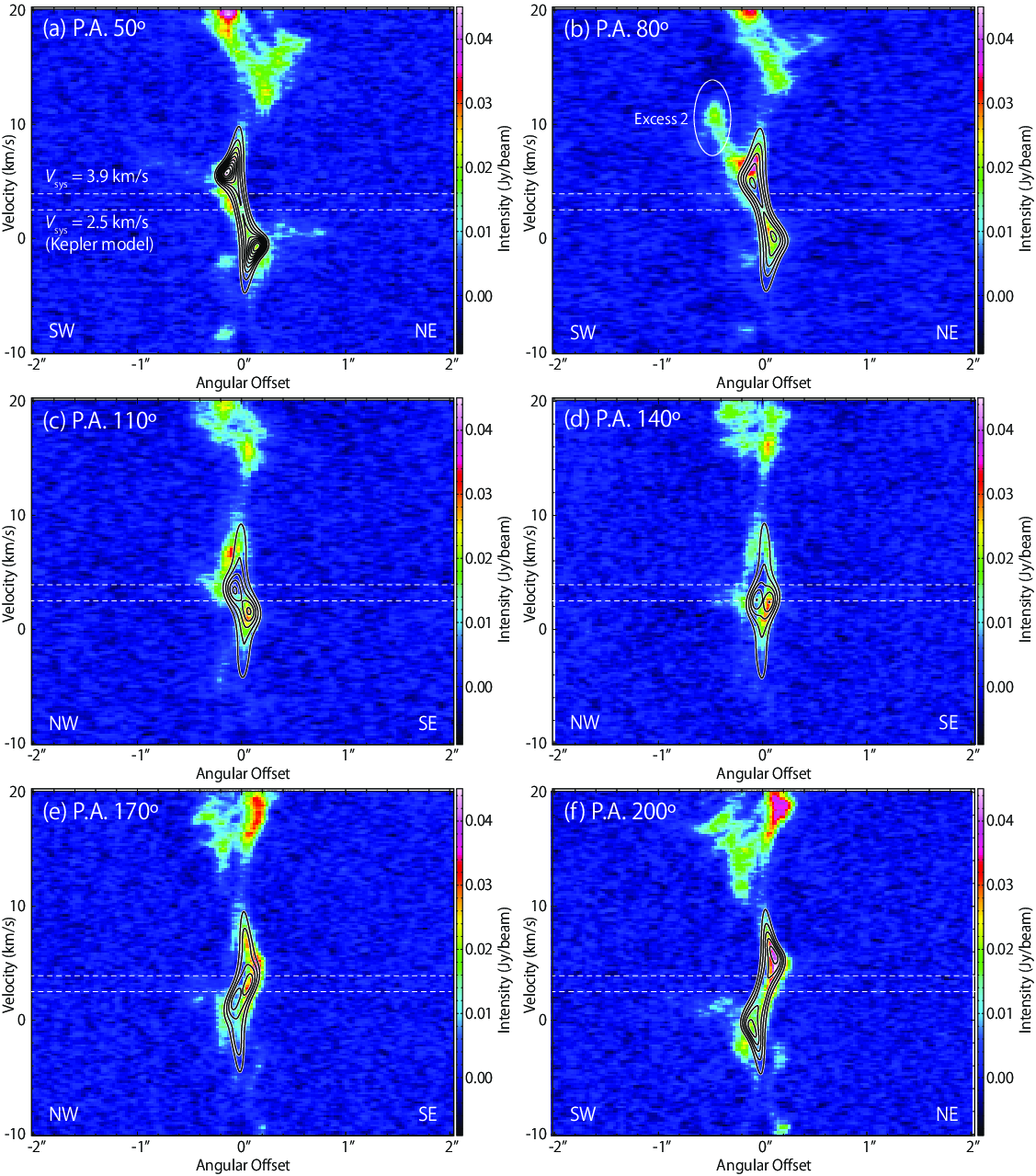}
	\fi
	\caption{Position-velocity diagrams of the \TFA\ (\tfad; color) line and the Keplerian disk model (contour). 
			Position axes in the panels are shown by arrows in {\bff Figure \ref{fig:PV_h2cs-707-725-744-Kep},} 
			which are centered at the continuum peak A1. 
			Black contours represent the same Keplerian model as in {\bff Figure \ref{fig:PV_h2cs-707-725-744-Kep}.} 
			{\bff The parameters employed for the model are as the same as those in Figures \ref{fig:PV_h2cs-707-725-744-Kep} and \ref{fig:PV_h2cs-744-Kep_MI-PA050}.} 
			Contour levels for the model result are at intervals of 10  \%\ of the peak intensity in the model cube. 
			\label{fig:PV_h2cs-744-Kep}} 
	\end{center}
\end{figure}

\begin{figure}
	\begin{center}
	\iffigurePV
	\vspace*{-20pt}
	\includegraphics[bb = 0 0 1000 1300, scale = 0.42]{\dirfig 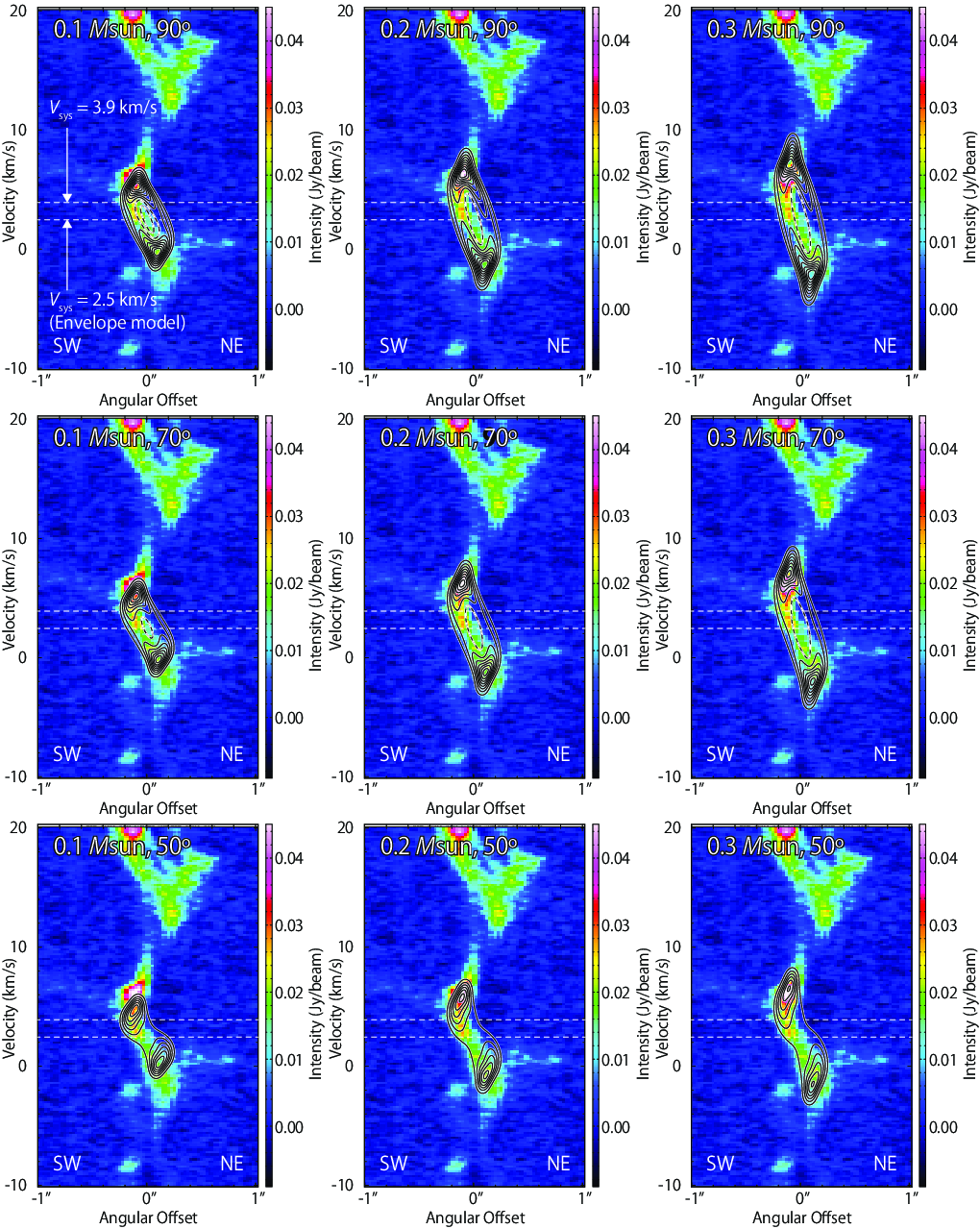}
	\vspace*{-10pt}
	\fi
	\caption{{\bff Same as Figure \ref{fig:PV_h2cs-744-Kep_MI-PA050}, 
			but for the \ire\ model.} 
			Black contours represent the \ire\ model results. 
			The physical parameters employed for the model are: 
			the central mass is 0.1, 0.2, and 0.3 \Msun, 
			and the \ia\ is 50\degr, 70\degr, and 90\degr. 
			The emissivity in the model is assumed to be proportional to $r^{-1.5}$, 
			where $r$ denotes the distance from the center of gravity, 
			and is to be zero for $r >$ \RoutireTFA. 
			The scale height of the disk is assumed to increase as increasing the distance from the center of gravity 
			with the flared angle of 30\degr. 
			Contour levels for the model result are at intervals of 10  \%\ of the peak intensity in the model cube. 
			{\bff The dashed contours in panels for the \ia\ of 80\degr\ and 90\degr\ represent the dip toward the central position.} 
			\label{fig:PV_h2cs-744-IRE_MI-PA050}} 
	\end{center}
\end{figure}

\begin{figure}
	\begin{center}
	\iffigurePV
	\vspace*{-20pt}
	\includegraphics[bb = 0 0 1000 1300, scale = 0.42]{\dirfig 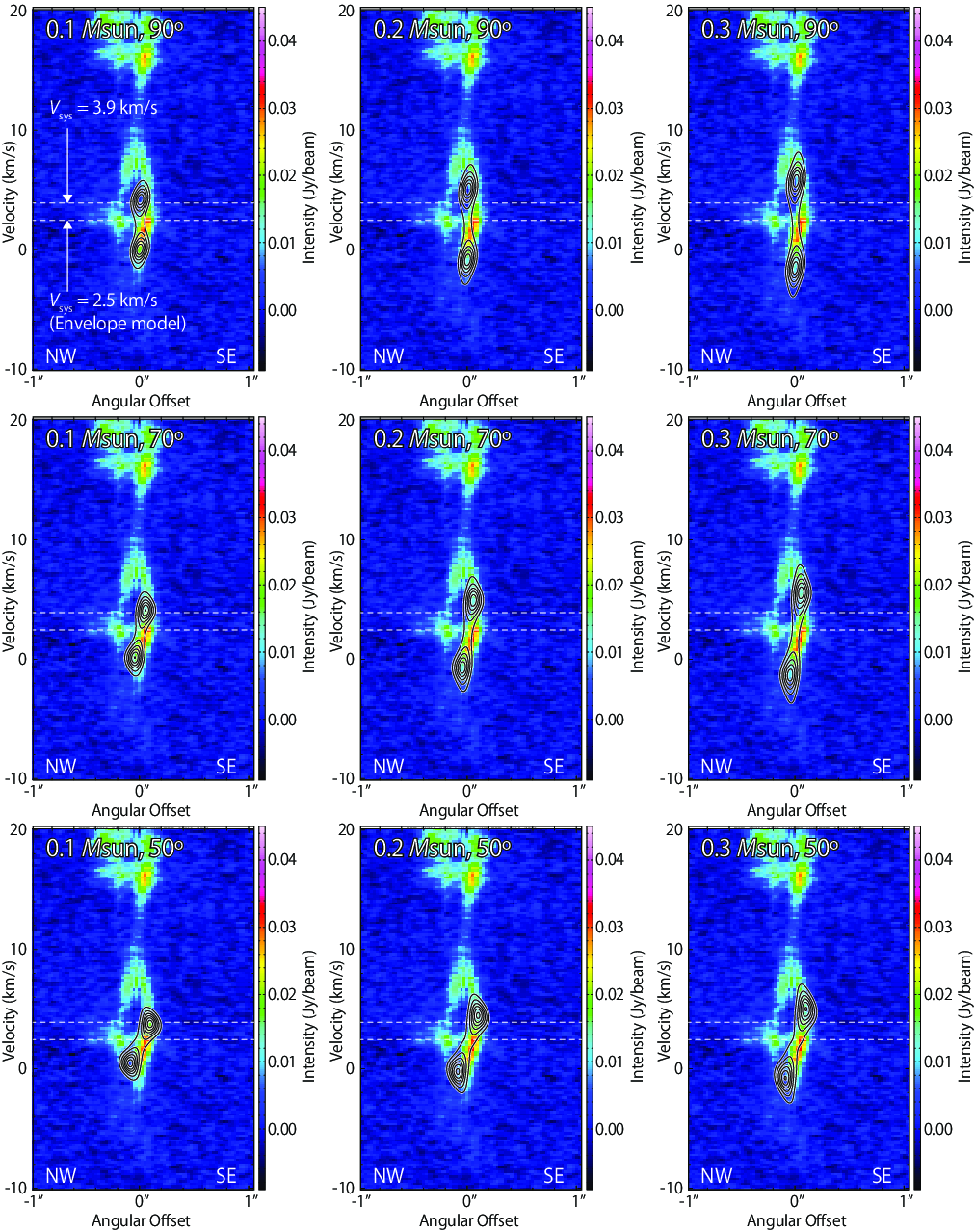}
	\vspace*{-10pt}
	\fi
	\caption{{\bff Same as Figure \ref{fig:PV_h2cs-744-IRE_MI-PA050}, 
			but along the direction perpendicular to the disk mid-plane direction (P.A. 140\degr).} 
			\label{fig:PV_h2cs-744-IRE_MI-PA140}} 
	\end{center}
\end{figure}

\begin{figure}
	\begin{center}
	\iffigurePV
	\includegraphics[bb = 0 0 1200 1250, scale = 0.4]{\dirfig 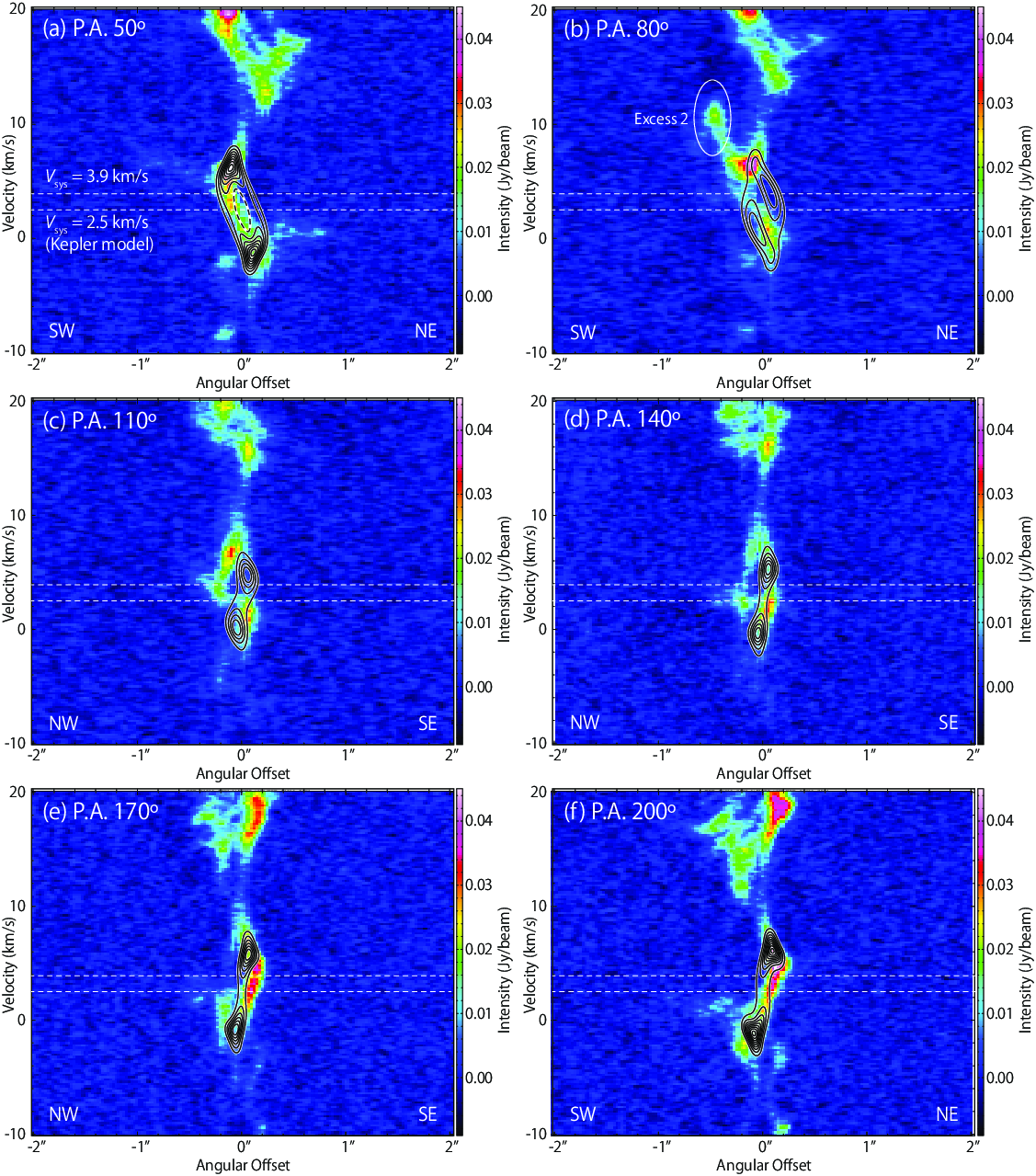}
	\fi
	\caption{{\bff Same as Figure \ref{fig:PV_h2cs-744-Kep}, 
			but for the \ire\ model.} 
			The physical parameters employed for the model are: 
			the central mass is \MireTFA, 
			the \ia\ is \IireTFA, 
			and the radius of the \cb\ is \CBireTFA. 
			{\bff Other details of the model are given in the caption of Figure \ref{fig:PV_h2cs-744-IRE}.} 
			Contour levels for the model result are at intervals of 10  \%\ of the peak intensity in the model cube. 
			The dashed contours in {\bff panel (a)} represent the dip toward the central position. 
			\label{fig:PV_h2cs-744-IRE}} 
	\end{center}
\end{figure}

\end{document}